\documentclass[12pt]{iopart}

\usepackage[scale=1.0]{OldStandard}
\usepackage{libertinust1math}
\usepackage{mylatexcommands}
\usepackage{qmdefs}
\usepackage{color}
\usepackage{graphicx}
\usepackage{ifthen}
\usepackage[]{cite}
\usepackage[]{mcite}

\newcommand{\figdir}{.}  

\newboolean{revision}
\setboolean{revision}{false}
\ifthenelse{\boolean{revision}}
{\newcommand{\authorcorr}{\color{red}}    \newcommand{\refereeone}{\color{blue}}   \newcommand{\refereetwo}{\color{magenta}}  \newcommand{\edteam}{\color{cyan}}}  
{\newcommand{\authorcorr}{\color{black}}  \newcommand{\refereeone}{\color{black}}  \newcommand{\refereetwo}{\color{black}}    \newcommand{\edteam}{\color{black}}}      

\definecolor{grey}{gray}{0.5}

\begin{document}

\newcommand{\gguide}{{\it Preparing graphics for IOP Publishing journals}}

\title[On Ehrenfest's theorem]{Considerations about the incompleteness of the Ehrenfest's theorem in quantum mechanics}

\author{Domenico Giordano}
\address{European Space Agency (retired), The Netherlands}  
\ead{dg.esa.retired@gmail.com}

\author{Pierluigi Amodio}
\address{Dipartimento di Matematica, Universit\`a di Bari, Italy}
\ead{pierluigi.amodio@uniba.it}

\vspace{10pt}
\begin{indented}
\item[]\today
\end{indented}

\begin{abstract}  We describe a study motivated by our interest to examine the incompleteness of the Ehrenfest's theorem in quantum mechanics and to resolve a doubt regarding whether or not the hermiticity of the hamiltonian operator is sufficient to justify a simplification of the expression of the macroscopic-observable time derivative that promotes the one usually found in quantum-mechanics textbooks.
The study develops by considering the simple quantum system ``particle in one-dimensional box''.
We propose theoretical arguments to support the incompleteness of the Ehrenfest's theorem in the formulation he gave, in agreement with similar findings already published by a few authors, and corroborate them with the numerical example of an electric charge in an electrostatic field.
{\edteam The contents of this study should be useful to Bachelor and Master students; the style of the discussions is tailored to stimulate, we hope, the student's ability for independent thinking.}  \end{abstract}

%
%
\noindent Published in the \textit{European Journal of Physics} on 16 September 2021 \\ Updated on 14 October 2021 \\
https://iopscience.iop.org/article/10.1088/1361-6404/abf69f \\
%
%
%


\section{Introduction\label{intro}}

In a short communication \cite{pe1927zfp} published in 1927, Ehrenfest showed that the macroscopic observables (also mean values or expectations values of) position $\mop$
and momentum $\mom$ of a point particle of mass $m$ are governed by Newtonian equations of motion {\authorcorr [Eqs. (4) and (5) at page 456]} under the action of the force 
\begin{equation} \label{pe.f}
  \mofpe
\end{equation}
imputable to the physical potential $V(x)$. 
He began from the time-dependent Schr\"{odinger} equation and its complex conjugate but did not give specific details about the proof; he just mentioned generic substitutions and integration by parts, and stressed that no {\authorcorr neglect} (\textit{ohne Vernachl\"{a}ssigung}) was needed.
One year later, in 1928, Ruark \cite{aer1928pr} extended Ehrenfest's theorem to a conservative system containing any number of particles.
The force on the generic particle, analog to \REq{pe.f}, in Ruark's procedure appeared accompanied by an additional term (equation at bottom of page 537) that was promptly dismissed by a synthetic declaration about its vanishing after integration by parts and subsequent application of Green's theorem; also Ruark did not give more specific details of his proof.
Ehrenfest's theorem is duly reported in quantum-mechanics textbooks \cite{am11961,cct1977, db1989,lb2000, pa2005,dg2005} and it has been positively valued by some authors \cite{dr1996ajp,cll2016arxiv} but scrutinised with critical eye by other authors{\authorcorr \cite{rnh1973ajp,nw1998me,va2000inc, va2001pla, sdv2013jop,sdv2013rbef}}.
Assuming, for the time being, the correctness of the theorem in the form
\begin{subequations} \label{pet}
	\begin{align}
	  \tds{}{\mop}{t} & = \frac{1}{m}\mom  \label{pet.p} \\[.5\baselineskip]
	  \tds{}{\mom}{t} & = \mofi \;,            \label{pet.m}
	\end{align}
\end{subequations}
the first aspect of \REqq{pet} targeted by criticism is the interpretation that they constitute Newtonian laws of motion.
Wheeler \cite{nw1998me} stressed vehemently the lack of rigour of such interpretation by pointing out that the equivalence
\begin{equation}\label{pot.eq}
   \mofi = - \left[\pd{}{V}{x} \right]_{x=\mop}
\end{equation}
necessary to give full physical meaning to the interpretation in question holds only under specific cases, the most renown of which is the harmonic oscillator.
In general, \REq{pot.eq} does not hold and the above set of differential equations [\REqq{pet}] is not in closed form from a mathematical point of view; a convincing example is provided by the one-dimensional Coulomb potential \mbox{$V(x)\sim 1/x$}.
Also Messiah \cite{am11961}, Jammer \cite{mj1989} and Ballentine \cite{lb2000} unambiguously pointed out the inappropriateness of the idea that the macroscopic observables $\mop$ and $\mom$ follow the laws of classical mechanics; moreover, Ballentine \etal \cite{lb1994pra} pushed the criticism even further despite the applicability of \REq{pot.eq}.
{\refereetwo Shankar \cite{rs1994} studied the equivalence of \REq{pet.m} with its classical-mechanics counterpart by carrying out an interesting analysis based on the concept of fluctuations and concluded by pointing out very clearly the quantitative limits of such an equivalence interpretation.}
Well, these critiques can be hardly argued against but, to some extent, they are mild.
Another criticism that, in our opinion, strikes harder and deeper is the one that questions the completeness of \REqq{pet}{\authorcorr; indeed, Hill \cite{rnh1973ajp}, Alonso \etal \cite{va2000inc, va2001pla} and De Vincenzo \cite{sdv2013jop,sdv2013rbef} detected additional terms appearing on their right-hand sides.} 
The purpose of our study is the investigation of this incompleteness matter parallel to the conceptual pathways traced by Hill, Alonso \etal, {\authorcorr and De Vincenzo} by bringing forth and elaborating on aspects that, to the best of our knowledge, do not seem to have been considered before.
{\refereeone 
We are going to look at the way boundary conditions are involved with respect to operator hermiticity, macroscopic-observable time derivative, and consequences on the Ehrenfest theorem, which is usually stated implicitly with the infinite-space physical domain in mind. 
The effect of three kinds of boundary conditions on the way that the theorem should be stated are considered and, for one of those kinds, we will corroborate the theoretical analysis with a significant numerical example.
Before launching ourselves onto the analysis' intricacies however, we find in order here to state clearly the following remark:
systematic exposition and discussion regarding the physical meaning of boundary conditions in quantum mechanics with respect to the generality of the subject matter is in no way meant to be the central theme of this work.
Good discussions about that theme were provided, for example, by Peierls \cite{rp1979}, Cassels \cite{jmc1982} and Reif \cite{fr2009}, although they did not specifically discuss boundary conditions in connection with Ehrenfest's theorem.}

\section{Macroscopic-observable time derivative and hermiticity\label{m-otd}}
The Ehrenfest's theorem surfaces in connection with the time derivative of the generic macroscopic observable $\moO$ associated with an operator $\Omega$ and generally defined \cite{pa2005} as
\begin{equation}\label{mo.def.g}
   \moO \cdot \int\wfc\,\wf\,d\tau = \int\wfc\Omega\wf d\tau 
\end{equation}
In \REq{mo.def.g}, $\tau$ is the {\authorcorr spatial} domain spanned by the coordinates required by the quantum system under consideration.
If it is possible to normalise the wavefunction
\begin{equation}\label{wfn}
   \int\wfc\,\wf\,d\tau = 1
\end{equation}
then \REq{mo.def.g} reduces to the standard form
\begin{equation}\label{mo.def}
   \moO = \int\wfc\Omega\wf d\tau
\end{equation}
A simple time-differentiation exercise \cite{db1989,pa2005,dg2005} that starts from \REq{mo.def} and takes advantage of the Schr\"{o}dinger equation
\begin{equation}\label{Seq}
  \ih \pd{}{\wf}{t} = \Ham \wf
\end{equation}
and its complex conjugate provides the expression
\begin{equation}\label{motd}
   \tds{}{\moO}{t} = \ioh \int \left[ \, \left(\Ham\wf\right)^{\ast} \left(\Omega\wf\right) - \wfc\,\Ham\left(\Omega\wf\right) \,\right] d\tau  
                   + \ioh \int \wfc \, \com{\Ham}{\Omega}\,\wf\,d\tau + \int \wfc \, \pd{}{\Omega}{t } \,\wf\,d\tau 
\end{equation}
At this point, the standard move in textbooks consists in the dismissal of the integral
\begin{equation}\label{Iomega}
   I_{\,\Omega} = \int \left[ \, \left(\Ham\wf\right)^{\ast} \left(\Omega\wf\right) - \wfc\,\Ham\left(\Omega\wf\right) \,\right] d\tau
\end{equation}
on the basis of the claimed hermiticity of the hamiltonian operator.
The Ehrenfest's theorem in the form of \REqq{pet} springs from the simplified \REq{motd} 
\begin{equation}\label{motd.s}
   \tds{}{\moO}{t} = \ioh \int \wfc \, \com{\Ham}{\Omega}\,\wf\,d\tau + \int \wfc \, \pd{}{\Omega}{t} \,\wf\,d\tau
\end{equation}
when it is particularised to position and momentum operators and after specification of the hamiltonian operator.
{\authorcorr \REqb{motd.s} is also evoked \cite{am11961,pa2005,dg2005} as supporting proof of the seemingly obvious statement of conservation: the macroscopic observable of a time-independent operator $(\pdt{}{\Omega}{t}=0)$ that commutes with the hamiltonian $(\com{\Ham}{\Omega}=0)$ does not change with time and is, therefore, conserved.}

Hermiticity is an extremely important feature in quantum mechanics because it ensures real values of the macroscopic observables and, thus, confers physical significance to them; a complex-valued macroscopic observable would not make sense from a physical point of view \cite{pd1967,db1989,pa2005,dg2005}. 
Such a lucid motivation for hermiticity is hardly disputable. But, in general, hermiticity comes with a fee. 
Let us consider the energy operator
\begin{equation}\label{enop}
   \Omega = \mathrm{E}= \ih \pd{}{}{t}
\end{equation}
for example; according to the general definition [\REq{mo.def.g}], the corresponding macroscopic observable is
\begin{equation}\label{mo.E}
   \mo{E} \cdot \int\wfc\,\wf d\tau = \int\wfc\ih\pd{}{\wf}{t} d\tau 
\end{equation}
and its complex conjugate reads
\begin{equation}\label{mo.E.cc}
   \mo{E}^{\ast} \cdot \int\wf\,\wfc d\tau = - \int\wf\ih\pd{}{\wfc}{t} d\tau 
\end{equation}
Subtraction of \REq{mo.E.cc} from \REq{mo.E} yields
\begin{equation}\label{mo.E-Ec}
   \mo{E} - \mo{E}^{\ast}  = \ih \tds{}{}{t} \left[ \ln \left(\int\wfc\,\wf d\tau \right) \right]
\end{equation}
\REqb{mo.E-Ec} indicates that the fee for the hermiticity of the energy operator is the normalisation [\REq{wfn}] of the wavefunction; as a matter of fact, any arbitrary constant on the right-hand side of \REq{wfn} would do but we can always scale the wavefunction in \REq{Seq} so that the constant becomes equal to 1.
Thus, normalisation comes more as an obligation rather than a choice{\authorcorr; w}e must be able to coerce the wavefunction into compliance with \REq{wfn} in order to expect real values of the macroscopically observable energy
\begin{equation}\label{mo.E.r}
   \mo{E} = \mo{E}^{\ast}
\end{equation} 
If one day we confront a problem within which we are not able to normalise the wavefunction then \REq{mo.E.r} {\authorcorr breaks down} and we should really ponder carefully about what physical meaning the problem we are dealing with has \cite{pd1967,db1989,dg2005}.
Hereinafter, we assume the validity of \REq{wfn}.

Let us consider the hamiltonian operator now. 
From the Schr\"{o}dinger equation [\REq{Seq}] and \REq{mo.E} we obtain
\begin{equation}\label{mo.H}
   \mo{H} = \mo{E}
\end{equation}
from which we deduce that also the hamiltonian must be hermitean
\begin{equation}\label{H.her}
   \mo{H}^{\ast} - \mo{H} = \int\left[ \, \left(\Ham\wf\right)^{\ast} \wf - \wfc\,\Ham\wf \,\right] d\tau = 0
\end{equation}
From our point of view, we look at \REq{H.her} not as a warranty of but rather as a test for hermiticity; in other words, when we are given an explicit hamiltonian then we should subject it to \REq{H.her} to find out whether or not it can be certified as hermitean.
It is rather straightforward to anticipate that the execution of such a test will bring forth the boundary conditions involving wavefunction and its derivatives and that they will play a fundamental role for the positive outcome of the test.

Let us assume that the test is passed and the hamiltonian is indeed hermitean. 
If we compare {\authorcorr the integrals of} \REqd{Iomega}{H.her} we notice that the latter's integrand involves only the wavefunction and its complex conjugate while the former's integrand involves also the action of the operator $\Omega$ on the wavefunction. 
Therefore, we may legitimately ask: is the hermiticity of the hamiltonian sufficient to make the integral $I_{\,\Omega}$  [\REq{Iomega}] vanish regardless of which operator $\Omega$ intervenes in it?
{\authorcorr Usually, q}uantum-mechanics textbooks \cite{db1989,pa2005,dg2005} give a decidedly affirmative answer: if the integral $I_{\,\Omega}$ vanishes for the unitary operator \mbox{$\Omega=1$}, and therefore the hamiltonian is hermitean, then it vanishes for any other operator $\Omega$. 
{\authorcorr Proofs are} provided, of course{\authorcorr, but their weaknesses have been nailed down, for example, by Alonso \etal \cite{va2000inc, va2001pla} and De Vincenzo \cite{sdv2013jop,sdv2013rbef} whose analyses take advantage of refined quantum-mechanical terminology, concepts and notation.
Here, however, we take the humbler attitude of the student with} a scientifically curious (or maybe skeptical) mind {\authorcorr who does not master yet the level of quantum-mechanical sophistication of the analyses mentioned a few lines ago and, leaning on her/his skill with integral calculus,} may decide to find out the answer by going through the direct execution of the integral $I_{\,\Omega}$; after all, carrying on along that course does not seem a terribly difficult task to accomplish with an explicitly declared and simple hamiltonian.
This is exactly what we plan to do in the next sections by considering the typical simple quantum system ``particle in one-dimensional box''.
{\authorcorr In passing, before concluding this section, we wish to point out that the doubt injected by the question about the hamiltonian-hermiticity sufficiency to dismiss the integral $I_{\,\Omega}$ imperils also the conservation statement mentioned in connection with \REq{motd.s}.}

\section{Particle in one-dimensional box \label{pib}}

\subsection{{\authorcorr Preliminary} considerations \label{prec}}

The particle lives on the $x$ axis and is subjected to the physical potential $V(x)$.
The {\authorcorr spatial} domain of interest is the segment $[-L,+L]$ and all the integrals in \Rse{m-otd} become one-dimensional
\begin{equation}\label{intx}
   \int (\cdots) \, d\tau \rightarrow \int_{-L}^{+L} (\cdots) \, dx
\end{equation}
The particle's hamiltonian is simply
\begin{equation}\label{H.pib}
   \Ham = - \hpib\pd{2}{}{x} + V(x)
\end{equation}
Hereinafter, we reserve the term hamiltonian exclusively to {\authorcorr indicate} \REq{H.pib}.
The quantum mechanics of the particle is governed by the Schr\"{o}dinger equation
\begin{equation}\label{Seq.pib}
  \ih \pd{}{\wf}{t} = - \hpib\pd{2}{\wf}{x} + V \wf
\end{equation}
whose integration calls for an initial condition
\begin{equation}\label{Seq.pib.ic}
  \wf(x,0) = f(x)
\end{equation}
and appropriate boundary conditions. 
We consider three sets of wavefunction boundary conditions
\begin{equation}\label{bc}
  \begin{cases}
      \wf(-L,t) = \wf(+L,t) = 0                                                       & \qquad \text{(c)}  \\[3ex]  
      \wf(-L,t) = \wf(+L,t) \quad ; \quad \pds{}{\wf}{x}{x=-L} = \pds{}{\wf}{x}{x=+L} & \qquad \text{(p)}     \\[3ex]
      \pds{}{\wf}{x}{x=-L} = \pds{}{\wf}{x}{x=+L} = 0                                 & \qquad \text{(v)}
  \end{cases}
\end{equation}
Initially confinement [\REqc{bc}{(c)}] attracted our main interest from a physical point of view but then, after becoming familiar with the papers of Hill \cite{rnh1973ajp} and Alonso \etal \cite{va2000inc,va2001pla}, we decided to incorporate in the investigation also periodicity [\REqc{bc}{(p)}] and vanishing-derivative [\REqc{bc}{(v)}] conditions in order to find out if we could retrieve the results already obtained by the mentioned authors.
We understand that the possibility to impose the periodic boundary conditions implies, or better is a consequence of, the periodicity of the potential
\begin{equation}\label{V.per}
   V(+L) = V(-L) 
\end{equation}
As it is well known, the general solution of \REq{Seq.pib} can be cast in the form of a series expansion
\begin{equation}\label{Seq.pib.gs}
   \wf(x,t) = \sum_{r} c_{r}\cdot \exp\left(- i \frac{\epsilon_{r} t}{\hbar}\right) \psir(x)
\end{equation}
whose coefficients are established by the initial wavefunction
\begin{equation}\label{Seq.pib.coe}
   c_{r} = \int_{-L}^{+L} \psi_{r}^{\ast}(x)\,f(x) dx
\end{equation}
In \REq{Seq.pib.gs}, $\epsilon_{r}$ and $\psir$ are eigenvalues and eigenfunctions of the time-independent Schr\"{o}dinger equation
\begin{equation}\label{tiSeq.pib}
   \Ham\psi = - \hpib\pd{2}{\psi}{x} + V\psi = \epsilon \psi
\end{equation}
which must be integrated with the eigenfunction boundary conditions
   \begin{equation}\label{tibc}
      \begin{cases}
          \psi(-L) = \psi(+L)=0                                                                                      & \qquad \text{(c)}  \\[3ex]  
          \psi(-L) = \psi(+L) \mbox{\hspace*{0.5em} ; \hspace*{0.5em}} \pds{}{\psi}{x}{x=-L} = \pds{}{\psi}{x}{x=+L} & \qquad \text{(p)} \\[3ex]
          \pds{}{\psi}{x}{x=-L} = \pds{}{\psi}{x}{x=+L}=0                                                            & \qquad \text{(v)}
      \end{cases}
   \end{equation}
in obvious compliance with the selection [\REqq{bc}] to be satisfied by the wavefunction.

\subsection{Hermiticity test for the hamiltonian\label{hht}}

The execution of the test requires the substitution of the hamiltonian in \REq{H.her} and the subsequent the integration; this series of operations leads to the following expression
\begin{equation}\label{H.pib.ht}
   \mo{H}^{\ast} - \mo{H} = \hpib \left[ \left( \wfc\wfd - \wf\wfcd \right)_{x=+L}  -  \left( \wfc\wfd - \wf\wfcd \right)_{x=-L}  \right]
\end{equation}
{\refereeone which is in accordance with (and, to some extent, even more formally consistent of) Eq. (4.3) at page 19 of Cassels' textbook \cite{jmc1982}.}
\REqb{H.pib.ht} shows explicitly how certification for hermiticity goes through the wavefunction boundary conditions.
As far as  our hamiltonian is concerned, we are on safe ground with the sets we are considering because the right-hand side of \REq{H.pib.ht} vanishes identically
\begin{equation}\label{H.pib.her}
   \mo{H}^{\ast} - \mo{H} = 0
\end{equation}
with each one of \REqq{bc} but such an occurrence should not give us a false sense of conviction that positive hermiticity certification is necessarily going to happen with other operators. 
For example, the momentum operator 
\begin{equation}\label{Mom}
   \Mom = \hoi \pd{}{}{x}
\end{equation}
is not so lucky
\begin{equation}\label{M.pib.ht}
   \mo{p}^{\ast} - \mo{p} = - \int_{-L}^{+L}\wf\hoi\wfcd \, dx \, - \int_{-L}^{+L}\wfc\hoi\wfd \, dx = \ih \left[ \left(\wf\wfc\right)_{x=+L} - \left(\wf\wfc\right)_{x=-L}   \right]
\end{equation}
{\refereeone \REqb{M.pib.ht} is a more formally consistent version of the analogous \mbox{Eq.~(4.2)} provided by Cassels at page 19 of his textbook \cite{jmc1982}.} 
The rightmost-hand side of \REq{M.pib.ht} survives, in general, with the vanishing-derivative boundary conditions and therefore the momentum operator is not hermitean in {\authorcorr the (v)} case. 
Such a state of affairs casts {\authorcorr obvious} doubts about the physical meaning of imposing vanishing wavefunction's derivatives at the boundaries{\authorcorr. H}owever, here we wish only to raise a warning flag but we will not go down that conceptual road any further because its ramifications deviate substantially from the main stream of our investigation's theme{\refereeone, as we have already pointed out clearly at the end of \Rse{intro}; Alonso \etal \cite{va2000inc,va2001pla} and De Vincenzo \cite{sdv2013rbef} provide good entry points to that conceptual road for the reader interested in pursuing the matter further.
Nevertheless, doubts notwithstanding, we do not consider the physical-meaning fragility of the (v) case an impelling reason for dismissal from our investigation yet; on the contrary, we believe that, at least for the time being, the (v) case contains an added pedagogical value because it exposes the student to an outstanding example 
that teaches an important lesson, to keep always in mind, about a recurring luring trap (particularly in quantum mechanics): even if the mathematics required by a physical problem flows smoothly, works finely (as we soon will see) and, in so doing, leaves us with an ecstatic sensation of respectful wonder, that is still insufficient warrant to confer physical solidity to results and conclusions.}

{\color{black}
Coming back to the hamiltonian, we can go one level deeper than \REq{H.pib.ht} if we insert the wavefunction general solution in \REq{H.her} but hang on with the replacement of the hamiltonian; if we do so, we obtain
\begin{equation}\label{H.pib.ht.gs}
   \mo{H}^{\ast} - \mo{H} = \sum_{r,s} c_{r}^{} c_{s}^{\ast} \exp\left(i \frac{\epsilon_{s}^{\ast} - \epsilon_{r}}{\hbar} t \right) 
                            \int_{-L}^{+L} \left( \psir\Ham \psisc - \psisc\Ham \psir \right) dx
\end{equation}
The integrals in \REq{H.pib.ht.gs} can be expanded in two different, although equivalent, ways.
The first one proceeds from the explicit substitution of the hamiltonian and subsequent execution of the integrals to yield
\begin{equation}\label{I.Hway}
    \int_{-L}^{+L} \left( \psir\Ham \psisc - \psisc\Ham \psir \right) dx  = 
   \hpib \left[ 
   \left( \psisc \pd{}{\psir}{x} - \psir \pd{}{\psisc}{x} \right)_{x=+L}  
   -  
   \left( \psisc \pd{}{\psir}{x} - \psir \pd{}{\psisc}{x} \right)_{x=-L}   \right]
\end{equation}
The second one  takes advantage of \REq{tiSeq.pib} and gives
\begin{equation}\label{I.Seqway}
   \int_{-L}^{+L} \left( \psir\Ham \psisc - \psisc\Ham \psir \right) dx = 
   \left( \epsilon_{s}^{\ast} - \epsilon_{r} \right) \int_{-L}^{+L} \psisc\psir dx
\end{equation}
The enforcement of any set in the eigenfunction boundary conditions, which, by the way, we recall are a direct consequence of the wavefunction boundary conditions whose enforcement sanctioned the hermiticity of the hamiltonian [\REqd{H.pib.ht}{H.pib.her}], makes vanish the right-hand side of \REq{I.Hway} and produces in cascade the following consequences.
\REqb{I.Hway} reduces to 
\begin{equation}\label{I.Hway.her}
    \int_{-L}^{+L} \left( \psir\Ham \psisc - \psisc\Ham \psir \right) dx  = 0
\end{equation}
which looks like the analog of \REq{H.her} in terms of the eigenfunctions and is taken as the starting point to introduce the concept of hermiticity in some textbooks \cite{pa2005}.
With \REq{I.Hway.her} in hand, \REq{H.pib.ht.gs} reconfirms the hermiticity of the hamiltonian.
\REqb{I.Seqway} simplifies to
\begin{equation}\label{I.Seqway.her}
   \left( \epsilon_{s}^{\ast} - \epsilon_{r} \right) \int_{-L}^{+L} \psisc\psir dx = 0
\end{equation}
\REqb{I.Seqway.her} is the basis to prove the reality of the eigenvalues and the orthogonality of the eigenfunctions when \mbox{$s\neq r$}.
If degenerate eigenfunctions turn out to exist then we can make recourse to the Gram-Schmidt recipe \cite{am11961,dg2005} to orthogonalise them but, for the sake of simplicity, we proceed with the assumption of degeneracy absence.
In any case, we are free to choose normalised eigenfunctions
\begin{equation}\label{efn}
   \int_{-L}^{+L} \psisc\psir dx = \delta_{sr}
\end{equation}
We have revised and collected \REqs{H.pib.ht.gs}{efn} here for convenience; {\authorcorr some of them} will be needed in the forthcoming sections.
}

\subsection{The integral $I_{\,\Omega}$\label{Iom}}

The substitution of the hamiltonian in \REq{Iomega} and the subsequent execution of the integral leads easily to
\begin{equation}\label{Iomega.pib}
   I_{\,\Omega} = \hpib  \left\{\left[ \wfc \pd{}{}{x}\left( \Omega\wf \right) - \left( \Omega\wf \right) \pd{}{\wfc}{x}\right]_{x=+L}
                -        \left[ \wfc \pd{}{}{x}\left( \Omega\wf \right) - \left( \Omega\wf \right) \pd{}{\wfc}{x}\right]_{x=-L} \right\}
\end{equation}
We see clearly from \REq{Iomega.pib} that the influence of the operator $\Omega$ on the fate of the integral $I_{\,\Omega}$ is unavoidable and not generalisable somehow; it seems that there is no evident manner to use the vanishing of the right-hand side of \REq{H.pib.ht} to imply the unconditional vanishing of the right-hand side of \REq{Iomega.pib} whatever the operator $\Omega$ is.
In other words, we are left with no other option than to assign explicitly the operator $\Omega$ and see what happens. 
Before engaging in such a task, however, {\authorcorr it is convenient at this point} to adapt the integral of the commutator in \REq{motd} to the hamiltonian
\begin{equation}\label{comm}
   I_{c} = \int_{-L}^{+L} \wfc \, \com{\Ham}{\Omega}\,\wf\, dx = - \hpib \int_{-L}^{+L} \wfc \, \com{\pd{2}{}{x}}{\Omega}\,\wf\, dx + \int_{-L}^{+L} \wfc \, \com{V}{\Omega}\,\wf\, dx
\end{equation}
The last integral in \REq{motd} plays no role because we will consider only time-independent operators.

\subsection{Position operator \label{po}} 
The substitution of the position operator \mbox{$\Omega=x$} in \REq{Iomega.pib} and further manipulation of the terms yields in general
\begin{equation}\label{Ix.pib}
    I_{\,\Omega=x} = \hpib  \left\{ 
                     \left[ \wfc\wf + x \left( \wfc\wfd - \wf\wfcd \right)  \right]_{x=+L}  -  \left[ \wfc\wf + x \left( \wfc\wfd - \wf\wfcd \right)  \right]_{x=-L}
                   \right\}
\end{equation}
or more {\authorcorr explicitly}
\begin{equation}\label{Ix.pib.bc}
    I_{\,\Omega=x} = \hpib \cdot
            \begin{cases}
                 0                                                                                                 & \qquad \text{(c)} \\[2ex]
                 2L \left[ \left(\wfc \pd{}{\wf}{x}\right)_{x=+L} - \left(\wf \pd{}{\wfc}{x}\right)_{x=-L} \right] & \qquad \text{(p)} \\[3ex]
                 \left( \wfc \wf \right)_{x=+L} - \left( \wfc\wf \right)_{x=-L}                                    & \qquad \text{(v)} \\[0ex]
            \end{cases} 
\end{equation}
by taking into account the wavefunction boundary conditions.
\REqb{Ix.pib.bc} shows flagrantly and unequivocally the evidence that the answer to the question we asked at the end of \Rse{m-otd} regarding the sufficiency of the hamiltonian's hermiticity to imply the unconditional vanishing of the integral $I_{\,\Omega}$ should be negative. 
Indeed, {\authorcorr \REq{H.pib.her}} tells us that the hamiltonian is hermitean regardless of which wavefunction boundary conditions are enforced but \REq{Ix.pib.bc} clearly says that the integral $I_{\,\Omega=x}$ does not vanish with the periodic and vanishing-derivative boundary conditions.
The substitution of the position operator $\Omega=x$ in \REq{comm} leads to
\begin{equation}\label{Ic-x}
   I_{c} = - \frac{\ih}{m} \mom
\end{equation}
We can now combine \REqd{Ix.pib.bc}{Ic-x} as required by \REq{motd}, and rearrange the terms a little bit for convenience, to obtain the time derivative of the macroscopic observable position
\begin{equation}\label{td-x}
   \tds{}{\mop}{t} = \frac{1}{m} \mom  +
                     \frac{\ih}{2m}\cdot
                                \begin{cases}
                                    0                                                                                                 & \qquad \text{(c)} \\[2ex]
                                    2L \left[ \left(\wfc \pd{}{\wf}{x}\right)_{x=+L} - \left(\wf \pd{}{\wfc}{x}\right)_{x=-L} \right] & \qquad \text{(p)} \\[3ex]
                                    \left( \wfc \wf \right)_{x=+L} - \left( \wfc\wf \right)_{x=-L}                                    & \qquad \text{(v)} \\[0ex]
                                \end{cases} 
\end{equation}
We deduce from \REq{td-x} that only confinement (c) produces an equation compliant with the {\authorcorr one} given by Ehrenfest [\REq{pet.p}]; the other boundary conditions do not.
With the replacement
\begin{equation}\label{hill}
   \left(\wf \pd{}{\wfc}{x}\right)_{x=-L} = \left(\wf \pd{}{\wfc}{x}\right)_{x=+L}
\end{equation}
permitted by the periodic boundary conditions, \REqc{td-x}{(p)} can be slightly modified into the form 
\begin{equation}\label{td-x-hill}
   \tds{}{\mop}{t} = \frac{1}{m} \mom  -
                     \frac{\ih}{2m}\cdot 2L \left[ \left(\wf \pd{}{\wfc}{x}\right)_{x=+L} - \left(\wfc \pd{}{\wf}{x}\right)_{x=+L} \right]
\end{equation}
that coincides with the equation given at the bottom of page 737 in Hill's paper \cite{rnh1973ajp} (notation correspondence: $2L\rightarrow (b-a)$; $\wfc\rightarrow\bar{\wf}$) and with \mbox{Eq.~(A.1)} at page 163 of Alonso \etal's paper \cite{va2000inc} (notation correspondence: $2L\rightarrow L$; $\wfc\rightarrow\bar{\wf}$).
We can go one level deeper than \REq{td-x} by substituting into it the wavefunction general solution to obtain the expression for the time derivative 
\begin{multline}\label{td-x.ef}
   \tds{}{\mop}{t} = \frac{1}{m} \mom \;  + \\
                     \frac{\ih}{2m}\cdot \sum_{r,s} c_{r}^{} c_{s}^{\ast}\exp\left(i \frac{\epsilon_{s} - \epsilon_{r}}{\hbar} t \right) \cdot
                                \begin{cases}
                                    0                                                                                                 & \text{(c)} \\[2ex]
                                    2L \left[ \left( \psisc \pd{}{\psir}{x}  \right)_{x=+L} - \left( \psir \pd{}{\psisc}{x}\right)_{x=-L} \right] & \text{(p)} \\[3ex]
                                    \left( \psisc \psir \right)_{x=+L} - \left( \psisc \psir \right)_{x=-L}   & \text{(v)} \\[0ex]
                                \end{cases} 
\end{multline}
in terms of the eigenfunctions.
{\authorcorr The motivation for} the procedure we have followed so far to deduce the time derivative of $\mop$ resides in the necessity to find out the behaviour of the integral $I_{\,\Omega}$ [\REq{Iomega.pib}] with respect to the established hermiticity of the hamiltonian.
However, there is an alternative{\authorcorr,} and certainly more direct{\authorcorr,} manner to deduce the time derivative in question; it starts from the definition of $\mop$ that takes advantage of the wavefunction general solution
\begin{equation}\label{mo.x}
   \mop = \int_{-L}^{+L} \wfc x \wf dx = \sum_{r,s} c_{r}^{} c_{s}^{\ast}\exp\left(i \frac{\epsilon_{s} - \epsilon_{r}}{\hbar} t \right) \cdot 
   \int_{-L}^{+L} \psisc x \psir dx
\end{equation}
A straightforward time derivation of \REq{mo.x} gives
\begin{equation}\label{td-x.ef.d}
   \tds{}{\mop}{t} = \ioh \sum_{r,s} c_{r}^{} c_{s}^{\ast} \left( \epsilon_{s} - \epsilon_{r} \right)  \exp\left(i \frac{\epsilon_{s} - \epsilon_{r}}{\hbar} t \right)  
      \cdot
      \int_{-L}^{+L} \psisc x \psir dx
\end{equation}
\REqb{td-x.ef.d} displays an interesting characteristic if compared to \REq{td-x.ef}: there is no explicit trace in it of any boundary conditions.
Of course, we expect equivalence between \REqd{td-x.ef}{td-x.ef.d} but we need to find out what is behind $\,\mom\,$ in order to have a better insight into this matter.
Therefore, we postpone the completion of this task to \Rse{tdmop}.

\subsection{Momentum operator \label{mo}}
We simply have to re-walk the algebraic trail followed for the position operator in {\authorcorr\Rse{po}}.
We substitute in \REq{Iomega.pib} the momentum operator \mbox{$\Omega=\Mom$} defined in \REq{Mom} and carry out the integral; a bit of attention should be payed to the appearance of the wavefunction's spatial second derivatives but they can be easily disposed of by extracting them from \REq{Seq.pib}.
When all the necessary algebra is taken care of, the final result reads
\begin{equation}\label{Ip.pib.bc}
    I_{\,\Omega=\Mom} = \hoi \cdot
            \begin{cases}
                 -\hpib \left[ \left( \pd{}{\wf}{x} \pd{}{\wfc}{x} \right)_{x=+L} - \left( \pd{}{\wf}{x} \pd{}{\wfc}{x} \right)_{x=-L}\right]  & \quad \text{(c)} \\[3ex]
                 0                                                                                                                             & \quad \text{(p)} \\[3ex]
                 \left( \wfc V \wf - \ih \wfc \pd{}{\wf}{t} \right)_{x=+L} - \left( \wfc V \wf - \ih \wfc \pd{}{\wf}{t} \right)_{x=-L}         & \quad \text{(v)} \\[0ex]
            \end{cases} 
\end{equation}
Here again we see that the integral may differ from zero according to the enforced boundary conditions; concerning confinement and periodicity, the situation in \REq{Ip.pib.bc} is reversed with respect to \REq{Ix.pib.bc}.
The substitution of the momentum operator \mbox{$\Omega=\Mom$} in \REq{comm} leads to the well known force term
\begin{equation}\label{Ic-p}
   I_{c} = \hoi \mofix
\end{equation}
The combination of \REqd{Ip.pib.bc}{Ic-p} according to \REq{motd} yields the time derivative of the macroscopic observable momentum
\begin{multline}\label{td-p}
   \tds{}{\mom}{t} = \mofix  \; + \\[2ex]
                     \begin{cases}
		                 -\hpib \left[ \left( \pd{}{\wf}{x} \pd{}{\wfc}{x} \right)_{x=+L} - \left( \pd{}{\wf}{x} \pd{}{\wfc}{x} \right)_{x=-L}\right]  &  \text{(c)} \\[3ex]
		                 0                                                                                                                             &  \text{(p)} \\[3ex]
		                 \left( \wfc V \wf - \ih \wfc \pd{}{\wf}{t} \right)_{x=+L} - \left( \wfc V \wf - \ih \wfc \pd{}{\wf}{t} \right)_{x=-L}         &  \text{(v)} \\[0ex]
                     \end{cases} 
\end{multline}
\REqb{td-p} tells us that, this time, only the periodic boundary conditions (p) give back an equation compliant with the {\authorcorr one} given by Ehrenfest [\REq{pet.m}]; surprisingly, confinement (c) does not, its duly conforming to \REq{pet.p} via \REqc{td-x}{(c)} notwithstanding.
\REqcb{td-p}{(c)} coincides with \mbox{Eq.~(A.2)} at page 164 of Alonso \etal's paper \cite{va2000inc} (notation correspondence: $2L\rightarrow L$; $\wfc\rightarrow\bar{\wf}$).
As we did in {\authorcorr \Rse{po}}, here also we can push the details one step deeper than \REq{td-p} by taking advantage of the wavefunction general solution to bring forth the eigenfunctions 
\begin{multline}\label{td-p.ef}
   \tds{}{\mom}{t} = \sum_{r,s} c_{r}^{} c_{s}^{\ast}\exp\left(i \frac{\epsilon_{s} - \epsilon_{r}}{\hbar} t \right) \cdot \\[2ex]
            \cdot \left[   \mofixef{s}{r} - 
           \begin{cases}
              \hpib \left[ \left( \pd{}{\psir}{x} \pd{}{\psisc}{x} \right)_{x=+L} - \left( \pd{}{\psir}{x} \pd{}{\psisc}{x} \right)_{x=-L}\right]  &  \text{(c)} \\[3ex]
              0                                                                                                                             &  \text{(p)} \\[3ex]
              \left[ (\epsilon_{r} - V) \psisc \psir \right]_{x=+L} - \left[ (\epsilon_{r} - V) \psisc \psir \right]_{x=-L}         &  \text{(v)} \\[0ex]
           \end{cases}\right]    
\end{multline}
{\authorcorr A} more direct manner to deduce the time derivative of $\mom$ exists too.
It consists in carrying out the time derivative of the definition
\begin{equation}\label{mo.p}
   \mom = \int_{-L}^{+L} \wfc \hoi \pd{}{\wf}{x} dx = \sum_{r,s} c_{r}^{} c_{s}^{\ast}\exp\left(i \frac{\epsilon_{s} - \epsilon_{r}}{\hbar} t \right) \cdot 
   \int_{-L}^{+L} \psisc \hoi \pd{}{\psir}{x} dx
\end{equation}
to obtain an expression
\begin{equation}\label{td-p.ef.d}
   \tds{}{\mom}{t} = \sum_{r,s} c_{r}^{} c_{s}^{\ast} \left( \epsilon_{s} - \epsilon_{r} \right)  \exp\left(i \frac{\epsilon_{s} - \epsilon_{r}}{\hbar} t \right)  
      \cdot
      \int_{-L}^{+L} \psisc \pd{}{\psir}{x} dx
\end{equation}
once again apparently free from the presence of the boundary conditions.
\REqb{td-p.ef} looks rather complex because it involves explicitly integrals of the physical potential's derivative and boundary conditions while \REq{td-p.ef.d} looks relatively simpler; 
are they equivalent? 
The answer requires to expand the eigenstates' contributions
\begin{equation}\label{mo.F.ec}
   \mofixef{s}{r}
\end{equation}
appearing in \REq{td-p.ef} and we will come back to it in \Rse{tdmom}.
For the time being, we can {\authorcorr transform} \REq{td-x.ef} {\authorcorr a bit further} due to the availability of \REq{mo.p} and {\authorcorr complete} the task{\authorcorr,} put on hold at the end of {\authorcorr\Rse{po},} concerned with finding out whether or not \REqd{td-x.ef}{td-x.ef.d} are equivalent. 


\subsection{{\authorcorr Equivalence of alternative} expressions for the time derivative of the macroscopic observable position  \label{tdmop}}
The substitution of \REq{mo.p} into \REq{td-x.ef} yields
\begin{multline}\label{td-x.ef.c}
   \tds{}{\mop}{t} = \sum_{r,s} c_{r}^{} c_{s}^{\ast}\exp\left(i \frac{\epsilon_{s} - \epsilon_{r}}{\hbar} t \right) \cdot \\ \cdot
                     \left[ \frac{1}{m} \int_{-L}^{+L} \psisc \hoi \pd{}{\psir}{x} dx + \frac{\ih}{2m}
                            \begin{cases}
                                0                                                                                                 & \text{(c)} \\[2ex]
                                2L \left[ \left( \psisc \pd{}{\psir}{x}  \right)_{x=+L} - \left( \psir \pd{}{\psisc}{x}\right)_{x=-L} \right] & \text{(p)} \\[3ex]
                                \left( \psisc \psir \right)_{x=+L} - \left( \psisc \psir \right)_{x=-L}   & \text{(v)} \\[0ex]
                            \end{cases} 
                     \right]
\end{multline}
Visual inspection indicates that the equivalence proof requires the term
\begin{equation*}
   \ioh \left( \epsilon_{s} - \epsilon_{r} \right) \cdot \int_{-L}^{+L} \psisc x \psir dx
\end{equation*}
of \REq{td-x.ef.d} to coincide with the term
\begin{equation*}
   \frac{1}{m} \int_{-L}^{+L} \psisc \hoi \pd{}{\psir}{x} dx + \frac{\ih}{2m}
                               \begin{cases}
                                   0                                                                                                 & \text{(c)} \\[2ex]
                                   2L \left[ \left( \psisc \pd{}{\psir}{x}  \right)_{x=+L} - \left( \psir \pd{}{\psisc}{x}\right)_{x=-L} \right] & \text{(p)} \\[3ex]
                                   \left( \psisc \psir \right)_{x=+L} - \left( \psisc \psir \right)_{x=-L}   & \text{(v)} \\[0ex]
                               \end{cases}
\end{equation*}
of \REq{td-x.ef.c}.
In order to show that this is indeed what happens, we need to rewind to \REq{I.Seqway}.
As it is well known, the equation's integral form descends from the equality of the respective integrands
\begin{equation}\label{I.Seqway.i}
      \left( \epsilon_{s}^{\ast} - \epsilon_{r} \right) \psisc\psir = \left( \psir\Ham \psisc - \psisc\Ham \psir \right)       
\end{equation}
In \REq{I.Seqway.i}, we can remove the complex-conjugation symbol $^{\ast}$ from $\epsilon_{s}$ because the hamiltonian is hermitean; moreover, we can adapt the right-hand side with the help of \REq{I.Hway} reformulated in local form
\begin{equation}\label{I.Hway.loc}
    \psir\Ham \psisc - \psisc\Ham \psir = 
   \hpib  \pd{}{}{x}\left( \psisc \pd{}{\psir}{x} - \psir \pd{}{\psisc}{x} \right)  
\end{equation}
\REqb{I.Hway.loc} turns \REq{I.Seqway.i} into
\begin{subequations}\label{proof}
\begin{equation}\label{p.i}
      \left( \epsilon_{s} - \epsilon_{r} \right) \psisc\psir = \hpib \pd{}{}{x}\left( \psisc \pd{}{\psir}{x} - \psir \pd{}{\psisc}{x} \right)   
\end{equation}
\REqb{p.i} is the starting point of the proof we are seeking; it can also be reached, and perhaps in a cleaner way, from the appropriate manipulation of the time-independent Schr\"{o}dinger equation [\REq{tiSeq.pib}] and its complex conjugate.  
Next step consists in the multiplication by $x i/\hbar$ on both sides
\begin{equation}\label{p.ii}
      \ioh \left( \epsilon_{s} - \epsilon_{r} \right) \psisc x \psir = \frac{\ih}{2m} x \pd{}{}{x}\left( \psisc \pd{}{\psir}{x} - \psir \pd{}{\psisc}{x} \right)   
\end{equation}
followed by the incorporation of $x$ inside the derivative and subsequent transformations
\begin{equation}\label{p.iii}
      \ioh \left( \epsilon_{s} - \epsilon_{r} \right) \psisc x \psir = 
      \frac{1}{m} \psisc\hoi \pd{}{\psir}{x} + 
      \frac{\ih}{2m} \pd{}{}{x}  \left[ x \left( \psisc \pd{}{\psir}{x} - \psir \pd{}{\psisc}{x} \right) + \psisc\psir \right] 
\end{equation}
Finally, the integration of \REq{p.iii} gives the proof of equivalence
\begin{multline}\label{p.iv}
   \ioh \left( \epsilon_{s} - \epsilon_{r} \right) \cdot \int_{-L}^{+L} \psisc x \psir dx = \\
      \frac{1}{m} \int_{-L}^{+L} \psisc \hoi \pd{}{\psir}{x} dx + \frac{\ih}{2m}
                                  \begin{cases}
                                      0                                                                                                 & \text{(c)} \\[2ex]
                                      2L \left[ \left( \psisc \pd{}{\psir}{x}  \right)_{x=+L} - \left( \psir \pd{}{\psisc}{x}\right)_{x=-L} \right] & \text{(p)} \\[3ex]
                                      \left( \psisc \psir \right)_{x=+L} - \left( \psisc \psir \right)_{x=-L}   & \text{(v)} \\[0ex]
                                  \end{cases}
\end{multline}
between \REqd{td-x.ef}{td-x.ef.d}.
The striking features of the simple left-hand side of \REq{p.iv} are: a) it spares us the necessity to calculate the eigenfunctions' spatial derivative contained in the integrand of the first term on the right-hand side, a perhaps numerically convenient circumstance in the cases in which the eigenfunctions are not available in analytical form; b) it carries smoothly and implicitly the knowledge of the boundary conditions contained explicitly in the second term on the right-hand side.
In conclusion, we remark that the presence of the triple term in \REq{td-x.ef} is indispensable for the proof of equivalence with \REq{td-x.ef.d}.
Such an indispensability hopelessly crosses out the attribute of {\authorcorr completeness} from Ehrenfest's \REq{pet.p}.
\end{subequations}

\subsection{{\authorcorr Equivalence of alternative} expressions for the time derivative of the macroscopic observable momentum \label{tdmom}}

In this section, we return to the question formulated at the end of {\authorcorr \Rse{mo}} concerned with the equivalence of \REqd{td-p.ef}{td-p.ef.d}.
We shift, therefore, our attention to the macroscopic observable force
\begin{equation}\label{mo.F}
   \mofix = \sum_{r,s} c_{r}^{} c_{s}^{\ast}\exp\left(i \frac{\epsilon_{s} - \epsilon_{r}}{\hbar} t \right)  \mofixef{s}{r}
\end{equation}
and, in particular, to the eigenstates' contributions [\REq{mo.F.ec}] in the addenda on the right-hand side of \REq{mo.F}.
We look at these terms as channels through which the influence of the environment on the particle becomes manifest; they are in a kind of stand-by state awaiting activation upon explicit assignment of the physical input represented by the potential $V(x)$.
Yet, the eigenstates' contributions possess a very interesting property that does not require the {\authorcorr explicit} knowledge of the potential and that can be dug up from the time-independent Schr\"{o}dinger equation [\REq{tiSeq.pib}], if we {\authorcorr arm ourselves} with a bit of prudent dexterity and, above all, solid patience to deal with the necessary algebra.

We begin from the time-independent Schr\"{o}dinger equation for the eigenstate $r$ and the complex conjugate for the eigenstate $s$
\begin{subequations} \label{pr}
	\begin{align}
	  - \hpib\pd{2}{\psir}{x}        + V\psir        & = \epsilon_{r} \psir         \label{tiSeq.pib.r} \\[.5\baselineskip]
	  - \hpib\pd{2}{\psisc}{x} + V\psisc & = \epsilon_{s} \psisc  \label{tiSeq.pib.s}
	\end{align}
and then take the spatial derivative of both
	\begin{align}
	  - \hpib\pd{2}{}{x}\psirx        + \Vx \psir        + V\psirx       & = \epsilon_{r} \psirx         \label{tiSeq.pib.r.i} \\[.5\baselineskip]
	  - \hpib\pd{2}{}{x}\psisx        + \Vx \psisc + V\psisx       & = \epsilon_{s} \psisx         \label{tiSeq.pib.s.i}
	\end{align}
After, we multiply \REq{tiSeq.pib.r.i} by $\psisc$, \REq{tiSeq.pib.s.i} by $\psir$, add them and rearrange terms to obtain
\begin{equation}\label{tiSeq.pib.rs.i}
   - \hpib \left(\psisc \pd{2}{}{x}\psirx + \psir \pd{2}{}{x}\psisx \right)            
   + 2 \underline{\psisc \Vx \psir} 
   + V \psirsx =
     \epsilon_{r} \psisc \psirx + \epsilon_{s} \psir \psisx       
\end{equation}
The term underlined in \REq{tiSeq.pib.rs.i} represents our target. The transformation of \REq{tiSeq.pib.rs.i} to the form
\begin{equation}\label{tiSeq.pib.rs.ii}
   \psisc \left( -\Vx \right) \psir = \left( \epsilon_{s} - \epsilon_{r} \right) \psisc \psirx + 
   \pd{}{}{x}\left[ \left( \epsilon_{r} - V \right) \psisc \psir + \hpib \psisx \psirx \right]
\end{equation}
involves juggling with derivatives' expansion and regrouping; second derivatives appear in the course of the process but they can be disposed of with the aid of \REqd{tiSeq.pib.r}{tiSeq.pib.s}.
The final step is the integration of \Req{tiSeq.pib.rs.ii} to yield
\begin{multline}\label{tiSeq.pib.rs.iii}
   \mofixef{s}{r} = \left( \epsilon_{s} - \epsilon_{r} \right) \int_{-L}^{+L} \psisc \pd{}{\psir}{x} dx \; + \\[2ex]
              \begin{cases}
                 \hpib \left[ \left( \pd{}{\psir}{x} \pd{}{\psisc}{x} \right)_{x=+L} - \left( \pd{}{\psir}{x} \pd{}{\psisc}{x} \right)_{x=-L}\right]  &  \text{(c)} \\[3ex]
                 0                                                                                                                             &  \text{(p)} \\[3ex]
                 \left[ (\epsilon_{r} - V) \psisc \psir \right]_{x=+L} - \left[ (\epsilon_{r} - V) \psisc \psir \right]_{x=-L}                 &  \text{(v)} \\[0ex]
              \end{cases}
\end{multline}
\REqb{tiSeq.pib.rs.iii} is as remarkable as \REq{p.iv}; if we introduce it into \REq{td-p.ef} then the triple term cancels out and what remains is precisely \REq{td-p.ef.d}.
We have, thus, proven the equivalence of \REqd{td-p.ef}{td-p.ef.d}.
The presence of the triple term in \REq{td-p.ef}, and{\authorcorr, \textit{a fortiori},} of the triple term in \REq{td-p}, is essential for the proof and inescapably removes the {\authorcorr attribute of completeness} from Ehrenfest's \REq{pet.m}.
\end{subequations}
{\refereeone
\subsection{Postliminary considerations\label{postc}}
\subsubsection{A moment of reflection.}
Let us spend a moment to reflect about what we have done and achieved so far.
In the paragraph concluding \Rse{m-otd}, we questioned the sufficiency of the hamiltonian's hermiticity to make the integral $I_{\,\Omega}$  [\REq{Iomega}] vanish regardless of which operator $\Omega$ intervenes in it.
In \Rsed{po}{mo}, we found out the answer to be a resolute and unquestionable ``No, it is not!'' by executing the integral $I_{\,\Omega}$ with position and momentum operators.
Systematic examination brought to surface the fundamental role played by the wavefunction boundary conditions and led to the time derivatives of the corresponding macroscopic observables given in \REqd{td-x}{td-p}.
The gears of the mathematical machinery moved smoothly and consistently, surprisingly even in the case of the vanishing-derivative (v) boundary conditions for which we know we are walking on a physically shaky ground due to non-hermiticity of the momentum operator and the consequent complex-valuedness of its macroscopic observable [\REq{M.pib.ht}].
Now, \REqd{td-x}{td-p} differ from \REqq{pet} and suggest, at least in principle, a quantum-mechanical macroscopic description of the particle's statics and dynamics substantially different from the, supposedly classical-mechanical, description provided by \REqq{pet}.  
So, it comes natural to ask: what is the physical meaning of the additional boundary terms appearing in \REqd{td-x}{td-p} but missing in Ehrenfest's formulation of his theorem?
Obviously, the question sounds meaningful only for confinement (c) and periodicity (p) boundary conditions.
It is clearly of doubtful significance for vanishing-derivative (v) boundary conditions due to the complex-valuedness shortcoming affecting the macroscopic observable momentum; as a matter of fact, with the obtainment of \REqc{td-x}{(v)} and \REqc{td-p}{(v)} we have been allowed to go as far as mathematical dexterity permits but, now that we specifically step into a purely physical context, we must diligently suspend consideration of the (v) case and concentrate our attention on the (c) and (p) cases. 


\subsubsection{Physical meaning of the boundary terms.}
\REqb{mo.def} is a good starting point for the quest regarding the physical meaning of the boundary terms in \REqd{td-x}{td-p}.
We look at its left-hand side with the classical-mechanics perspective in mind and we see a macroscopically observable property that belongs to the particle, a physical system that we (are habituated to) conceive punctiform in nature and, therefore, occupying geometrical points inside (and outside of) the spatial domain.
The right-hand side reflects the quantum-mechanics viewpoint: the same property is distributed in the spatial domain with density $\wfc\Omega\wf$, the physical system may still be the particle but our habit to conceive it punctiform and localised in geometrical points is not really necessary anymore.
We focus on the quantum-mechanics viewpoint and, with anticipated expectation of clarification from the fluid-dynamics formulation of quantum mechanics, pioneered by de Broglie \cite{ldb1926cras,ldb1927crasjj,ldb1927crasjd}, Madelung \cite{em1927zfp}, Bohm \cite{db1952pr166,db1952pr180}, Takabayasi \cite{tt1952ptp,tt1953ptp} 
and still being explored \cite{hew1970prd,tt1983ptp,pv2016f} in more recent years, we ask: how does the macroscopic-observable density $\wfc\Omega\wf$ change with time?
Well, we simply have to repeat the same time-differentiation exercise that leads to \REq{motd} but we have to work locally this time, and with the hamiltonian specified in \REq{H.pib}.
The reward at the end of the algebraic procedure is
\begin{subequations}\label{modtd.fdbe}
\begin{multline}\label{modtd}
   \pd{}{}{t}(\wfc\Omega\wf) + \pd{}{}{x} \left\{ \hotmi \left[ \wfc \pd{}{}{x}\left( \Omega\wf \right) - \left( \Omega\wf \right) \pd{}{\wfc}{x} \right] \right\} = \\
   \ioh\wfc\com{ \left(-\hpib \pd{2}{}{x} + V \right) }{\Omega}\wf + \wfc \, \pd{}{\Omega}{t } \,\wf
\end{multline} 
It is important to keep in mind that there is no physical information in \REq{modtd} that is not already contained in \REq{motd}, conjoined with \REq{H.pib}, and viceversa; both equations convey the same physics in local and integral forms, respectively.
\REq{modtd} reveals the typical mathematical structure of a fluid-dynamics balance equation; 
with regard to the macroscopically observable property $\moO$, there are: on the left-hand side, time derivative of its density and divergence of its flux
\begin{equation}\label{flux}
   S_{\,\Omega} =  \hotmi \left[ \wfc \pd{}{}{x}\left( \Omega\wf \right) - \left( \Omega\wf \right) \pd{}{\wfc}{x} \right] 
\end{equation}
responsible for the local transport throughout the spatial domain, and on the right-hand side, its production term
\begin{equation}\label{prod}
   P_{\Omega} =  \ioh\wfc\com{ \left(-\hpib \pd{2}{}{x} + V \right) }{\Omega}\wf + \wfc \, \pd{}{\Omega}{t } \,\wf
\end{equation}
that accounts for the creation/destruction per unit time in the unit length. 
In the last sentence, we have clearly spoken in fluid-dynamics parlance.
\end{subequations}
In order to be sure that we are on the right track, we set $\Omega=1$ in \REqd{modtd}{flux} to retrieve as particular case the equations for probability 
\begin{subequations}\label{modtd.fdbe.bohm}
  \begin{equation}\label{modtd.bohm}
	   \pd{}{}{t}(\wfc\wf) + \pd{}{}{x} \left[ \hotmi \left( \wfc \pd{}{\wf}{x} - \wf \pd{}{\wfc}{x} \right) \right] = 0
  \end{equation}
  and for its current density 
  \begin{equation}\label{flux.bohm}
     S_{\,\Omega=1} =  \hotmi \left( \wfc \pd{}{\wf}{x} - \wf \pd{}{\wfc}{x} \right) 
  \end{equation}
  that Bohm proposed at page 83 of his textbook \cite{db1989}.
  Additionally, \REq{prod} indicates that the production term vanishes identically
  \begin{equation}\label{prod.bohm}
     P_{\Omega=1} =  0
  \end{equation}
  and, in so doing, confers the conservation attribute to probability: it cannot be either created or destroyed, exactly as it happens for total mass in fluid dynamics.
  As a matter of fact, if we multiply \REqd{modtd.bohm}{flux.bohm} by $m$ then what we obtain may very well be interpreted as balance equation and flux of particle's mass.  
\end{subequations}
Bohm remarked wittingly
\begin{quote}
   This idea that probability flows through space more or less like a fluid is very useful physically.
\end{quote}
Yes, definitely. 
But probability is not the only one that ``flows through space'': \REq{modtd} unambiguously tells us that the same conceptual idea applies to any macroscopic observable.
If we integrate it on the spatial domain then the flux-divergence term will yield the difference of the fluxes evaluated at the boundaries
\begin{multline}\label{key}
   S_{\,\Omega}(+L) - S_{\,\Omega}(-L) = \\ \hotmi \left[ \wfc \pd{}{}{x}\left( \Omega\wf \right) - \left( \Omega\wf \right) \pd{}{\wfc}{x} \right]_{x=+L}
                                       -    \hotmi \left[ \wfc \pd{}{}{x}\left( \Omega\wf \right) - \left( \Omega\wf \right) \pd{}{\wfc}{x} \right]_{x=-L} 
                                       = \\ \parbox{0em}{\hspace*{-2em}$\dfrac{1}{\ih}\,I_{\Omega}$}
\end{multline}
which, save for the irrelevant multiplicative factor $1/\ih$, matches precisely the integral $I_{\,\Omega}$ [\REq{Iomega.pib}] that originates the boundary terms in \REqd{td-x}{td-p}. 
There is, therefore, only one possible physical interpretation for those terms consistent with the fluid-dynamics perspective to which \REq{modtd} belongs: they represents the exchange of the corresponding macroscopically observable property between the physical system residing inside the spatial domain and its external environment residing outside. 
Such exchange is necessarily governed by the wavefunction's boundary conditions and the fine details of its physical interpretation are achievable only after the specification of the operator $\Omega$.

\subsubsection{Position operator}
When dealing with the position operator, the application of \REq{modtd} should not be thoughtlessly rushed but meditated with a grain of circumspection.
Fluid dynamicists know very well that the physical meaningfulness of a balance equation hangs on an important requirement for the property being balanced: it must be extensive.
In other words, a balance equation like \REq{modtd} can be formulated only for properties of a physical system that are proportional to the extension of the spatial domain that contains the physical system \cite{ln1971}.
Mass, momentum, angular momentum, and energy are examples of extensive properties. 
Position is not but we can elude this minor conceptual obstacle by setting $\Omega=mx$ in \REqq{modtd.fdbe} in order to be formally compliant with the extensiveness requirement.  
Then we obtain the balance equation
\begin{subequations}\label{modtd.fdbe.x}
\begin{equation}\label{modtd.x}
   \pd{}{}{t}(\wfc mx\wf) + \pd{}{}{x} \left(S_{mx}\right) = P_{mx} 
\end{equation}
with flux 
\begin{equation}\label{flux.x}
   S_{mx} =  \hoit \left[ \wfc\wf + x \left( \wfc \pd{}{\wf}{x} - \wf \pd{}{\wfc}{x} \right) \right]
\end{equation}
and production
\begin{equation}\label{prod.x}
   P_{mx} =  \wfc\hoi\pd{}{\wf}{x}
\end{equation}
\end{subequations}
\REqb{modtd.x} constitutes the local form of \REq{td-x}.
We recognise from \REq{flux.x} the explicit correspondence between the boundary terms in \REq{td-x}, multiplied by $m$, and the exchange term generated by the formal integration of \REq{modtd.x} on the spatial domain
\begin{equation}\label{et.mx}
   S_{mx}(+L)-S_{mx}(-L) = - \frac{\ih}{2}\cdot
                                   \begin{cases}
                                       0                                                                                                 & \qquad \text{(c)} \\[2ex]
                                       2L \left[ \left(\wfc \pd{}{\wf}{x}\right)_{x=+L} - \left(\wf \pd{}{\wfc}{x}\right)_{x=-L} \right] & \qquad \text{(p)} \\[0ex]
                                   \end{cases}    
\end{equation}
We may also look at them as an offset between the macroscopically observable momentum $\mom$ of the particle and the amount of momentum attributable to the product of its mass and velocity \mbox{$m\cdot\tdt{}{\mop}{t}$}.
Thus, quantum mechanics and classical mechanics clearly agree in the (c) case because the exchange term vanishes but not in the (p) case because the exchange term stands firmly out.
In this regard, Hill \cite{rnh1973ajp} has investigated the statics of a free particle [$V(x)=0$] in a generic stationary state ($\tdt{}{\mop}{t}=0$) with periodic (p) boundary conditions ($\mom\neq0$) and has proven analytically the necessity of the additional momentum supplied by the exchange term to keep \REq{td-x-hill} in balance.
However, his explanation for such an occurrence, provided at page 738 of his article, is only of mathematical nature.
Also, Reif proposed at page 356 of his textbook \cite{fr2009} an interesting problem regarding the motion of a particle on a circumference, again involving periodic (p) boundary conditions therefore, that offers an appropriate scenario to scrutinise the physical meaning of the boundary terms in \REqc{td-x}{(p)}; De Vincenzo \cite{sdv2014rbef} has discussed classical and quantum descriptions of this problem in details and we refer the interested reader to his well-written paper. 

\subsubsection{Momentum operator}
If we substitute in \REqq{modtd.fdbe} the momentum operator \mbox{$\Omega=\Mom$} defined in \REq{Mom} then we obtain the particle's momentum balance equation
\begin{subequations}\label{modtd.fdbe.p}
\begin{equation}\label{modtd.p}
   \pd{}{}{t} \left( \wfc\hoi\pd{}{\wf}{x} \right) + \pd{}{}{x} \left(S_{\,\Mom}\right) = P_{\Mom} 
\end{equation}
with flux 
\begin{equation}\label{flux.p}
   S_{\,\Mom} =  \hpib \left( \pd{}{\wf}{x} \pd{}{\wfc}{x} - \wfc\pd{2}{\wf}{x} \right)
\end{equation}
and production
\begin{equation}\label{prod.p}
   P_{\Mom} =  \wfc \left( - \pd{}{V}{x} \right) \wf
\end{equation}
\end{subequations}
\REqqb{modtd.fdbe.p} display an outstanding analogy with fluid dynamics, or, perhaps better, with the mechanics of continuous media, that encourages to draw the following remarkable deductions.
In the spatial domain, there exists a stress distribution with tensor $-S_{\,\Mom}$ (there is only the $xx$-component in our one-dimensional case) and force density $-\pdt{}{S_{\,\Mom}}{x}$ induced by external forces applied at the boundaries that turn out to be precisely the boundary terms in \REq{td-p}
\begin{multline}\label{et.p}
   [-S_{\,\Mom}(+L)] - [-S_{\,\Mom}(-L)] =  \int_{-L}^{+L} \pd{}{}{x} \left(-S_{\,\Mom}\right)\, dx = \\
                                  \begin{cases}
			   		                 -\hpib \left[ \left( \pd{}{\wf}{x} \pd{}{\wfc}{x} \right)_{x=+L} - \left( \pd{}{\wf}{x} \pd{}{\wfc}{x} \right)_{x=-L}\right]  &  \text{(c)} \\[3ex]
			   		                 0                                                                                                                             &  \text{(p)} \\[0ex]
			                      \end{cases} 
\end{multline}
The external forces, also interpretable as exchange of momentum, are necessarily due to the confining walls in the (c) case and \REqc{et.p}{(c)}, perhaps surprisingly, tells us that those forces are finite, an outcome in full contraposition with the widespread belief that forces due to confining walls should be infinite instead.
The external forces vanish in the (p) case. 
Furthermore, the force density due to the physical potential [\REq{prod.p}] plays the role of a field responsible for the production of momentum.
For the sake of completeness, we wish to mention that both \REqd{modtd.p}{flux.p} can be turned into more mathematically symmetric, although clearly physically equivalent, forms.
The key for the transformation is the hermiticity of the momentum operator which holds in the (c) and (p) cases [\REq{M.pib.ht}]. 
The recipe for the transformation calls for the complex conjugation of \REq{modtd.p}, adding the outcome to \REq{modtd.p} and dividing by 2; that leads to the final balance equation
\begin{subequations}\label{modtd.fdbe.p.1}
\begin{equation}\label{modtd.p.1}
   \pd{}{}{t} \left[ \hoit  \left( \wfc\pd{}{\wf}{x} - \wf\pd{}{\wfc}{x} \right) \right] + \pd{}{}{x} \left(S_{\,\Mom}\,\!\!\!'\right) = P_{\Mom} 
\end{equation}
with flux
\begin{equation}\label{flux.p.1}
   S_{\,\Mom}\,\!\!\!' =  \hpib \left( \pd{}{\wf}{x} \pd{}{\wfc}{x} - \frac{1}{2}\wfc\pd{2}{\wf}{x}  - \frac{1}{2}\wf\pd{2}{\wfc}{x}\right)
\end{equation}
\end{subequations}
The momentum densities in \REqd{modtd.p}{modtd.p.1} are equivalent
\begin{equation}\label{m.dens}
   \mo{p} = \int_{-L}^{+L}\wf\hoi\wfcd \, dx = \mo{p}^{\ast} = - \hoi  \int_{-L}^{+L} \wf\pd{}{\wfc}{x} \, dx 
          = \hoit  \int_{-L}^{+L} \left( \wfc\pd{}{\wf}{x} - \wf\pd{}{\wfc}{x} \right) \, dx
\end{equation}
and the boundary-flux difference is expectedly invariant
\begin{equation}\label{bf.diff}
  S_{\,\Mom}(+L) - S_{\,\Mom}(-L) =  S_{\,\Mom}\,\!\!\!'(+L) - S_{\,\Mom}\,\!\!\!'(-L)
\end{equation}

\REqdb{modtd.p}{modtd.p.1} constitute the local form of \REq{td-p}.
By looking at them in an attempt to rescue Ehrenfest's formulation [\REqq{pet}] of his theorem at least in the (c) case, a concerned student may raise the following, admittedly rather gripping, question: is there a way to transform their flux-divergence term, induced by the external forces due to the confining walls, into a production term structured as \REq{prod.p}?
In mathematical terms, and selecting \REq{flux.p.1} for example, we should seek a potential $V'$ that satisfies the condition
\begin{subequations}\label{fd.to.p}
\begin{equation}\label{fd.to.p.a}
   \wfc \left( - \pd{}{V'}{x} \right) \wf = - \hpib \pd{}{}{x} \left( \pd{}{\wf}{x} \pd{}{\wfc}{x} - \frac{1}{2}\wfc\pd{2}{\wf}{x}  - \frac{1}{2}\wf\pd{2}{\wfc}{x}\right) 
\end{equation}
so that \REqc{td-p}{(c)} would undoubtedly turn into
\begin{equation}\label{fd.to.p.e}
    \tds{}{\mom}{t} = \int_{-L}^{+L} \mbox{\hspace*{-0.5em}}\wfc \left[ - \pd{}{}{x} \left( V + V' \right) \right] \wf dx  
\end{equation}
and justify the conclusion that if we consider the \textit{total} potential $(V+V')$ then Ehrenfest's formulation [\REq{pet}] of his theorem is fine because \textit{all} the forces are taken into account as derivable from a potential.
Indeed, it is relatively not difficult to extract the potential $V'$ from \REq{fd.to.p.a}.
As a matter of fact, we can even ignore the terms with the second derivatives because, upon integration, they vanish in the (c) case; thus, we are allowed to simplify the basic condition for the quest of the potential $V'$ as follows
\begin{equation}\label{fd.to.p.c}
	\wfc \left( - \pd{}{V'}{x} \right) \wf =  - \hpib \pd{}{}{x} \left( \pd{}{\wf}{x} \pd{}{\wfc}{x} \right) 
\end{equation}
In the (c) case, \REq{fd.to.p.c} predicts  
\begin{equation}\label{cwf}
   \left|- \pds{}{V'}{x}{x=\pm L}\right| \rightarrow \infty
\end{equation}
at the confining walls if we assume the finiteness of the right-hand side.
\end{subequations}
Admittedly, all appears correct mathematically. 
However, in our opinion, we believe there is no much physical content in the mathematical sophistry of \REqq{fd.to.p} because the potential $V'$ 
(i)   does not belong to the hamiltonian and, therefore, does not participate in the determination of the wavefunction $\wf$, 
(ii)  can be evaluated \textit{only after} that the wavefunction $\wf$ has been determined, and 
(iii) turns out to be, after all, just an \textit{ad hoc} term introduced to camouflage a flux divergence as a production for the purpose of making \REqc{td-p}{(c)} look like \REq{pet.m}.
Another attemptable rescue line could ignore \REqd{fd.to.p.a}{fd.to.p.c}, \textit{de facto} the boundary terms in \REqc{td-p}{(c)}, trust the unconditional applicability of both \REqd{fd.to.p.e}{cwf}, and blame the lack of operativeness of \REq{fd.to.p.e} to the drawback of the wall-related potential $V'$ emanating from \REq{cwf}: the force $-\pdt{}{V'}{x}$ cannot be cast in a form for which \REq{fd.to.p.e} can be directly applied.
This argumentation saves the form of \REq{fd.to.p.e} to the detriment of its operativeness; indeed, how can we then calculate the macroscopically observable dynamics of the particle in the (c) case if the operativeness of \REq{fd.to.p.e} is disabled by the presence of the confining walls?
Additionally, the reconciliation of \REq{fd.to.p.e} with the presence of the boundary terms in \REqc{td-p}{(c)} would still remain an intricate unsettled question.
When contemplating these rescue attempts, a witty statement of Sutcliffe (at top of page 35 of) \cite{bs1975} comes to mind to portray them as typical examples in which
\begin{quote}
... one artificially forces on the system one's preconceptions about behaviour.
\end{quote}

In conclusion, we feel comfortable with the physical interpretation attached to \REqq{modtd.fdbe.p} in local form and to \REqc{td-p}{(c,p)} in integral form because it is more complete, free of conceptual doubts and operative from a calculation point of view.

}

\subsection{Numerical test case \label{ntc}}

\subsubsection{{\authorcorr Introductory} reflections.\label{ic}}

The theoretical considerations of the {\authorcorr precedent} sections align consistently with those of {\authorcorr Hill \cite{rnh1973ajp}, Alonso \etal \cite{va2000inc, va2001pla} and De Vincenzo \cite{sdv2013jop,sdv2013rbef} regarding the presence of the boundary terms in \REqd{td-x}{td-p}} and invalidate the completeness of Ehrenfest's theorem in the formulation he gave [\REqq{pet}].
We wish to {\authorcorr complement and to} corroborate the theory {\refereeone dealt with so far} with a numerical {\refereeone test case because we are convinced that, from a pedagogical point of view, fully worked-out numerical examples help greatly the students to bring theoretical and mathematical abstraction within their grasp.
We believe it is prudent to follow Hill's choice \cite{rnh1973ajp} and start with a statics example.
Of course, a dynamics case is certainly more attractive because richer in details but we prefer to proceed in steps of increasing difficulty and, in this respect, we point the impatient reader to De Vincenzo's study \cite{sdv2013jop} of the one-dimensional dynamics of a particle subjected only to the external forces due to the confining walls $[V(x)=0]$.
}
We will consider only the confinement boundary conditions [\REqc{bc}{(c)} and \REqc{tibc}{(c)}]{\authorcorr; f}or convenience, we collect here the relevant equations [\REqc{td-x}{(c)} and \REqc{td-p}{(c)}]
\begin{subequations} \label{et.c}
	\begin{align}
	  \tds{}{\mop}{t} & = \frac{1}{m}\mom  \label{etc.p} \\[.5\baselineskip]
	  \tds{}{\mom}{t} & = \mofix - \sigma\cdot\hpib \left[ \left( \pd{}{\wf}{x} \pd{}{\wfc}{x} \right)_{x=+L} - \left( \pd{}{\wf}{x} \pd{}{\wfc}{x} \right)_{x=-L}\right]    \label{etc.m}
	\end{align}
\end{subequations}
In \REq{etc.m}, we have introduced the numerical switch
\begin{equation}\label{ns}
   \sigma = \begin{cases} 0 & \text{Ehrenfest \cite{pe1927zfp}} \\[1.5ex] 1 & \text{Alonso \etal \cite{va2000inc}, {\authorcorr De Vincenzo \cite{sdv2013jop,sdv2013rbef},} this study} \end{cases}
\end{equation}
to be able to select between the formulations whose consequences we wish to compare.
Of course, \REqd{td-x.ef.d}{td-p.ef.d} remain applicable as such.
We can dress {\authorcorr the external forces due to the confining walls in \REq{etc.m}} in a perhaps more physically significant manner by applying the transformation
\begin{equation}\label{bct}
    \pd{}{\wf}{x} \pd{}{\wfc}{x} = \frac{1}{2} \pd{2}{(\wfc\wf)}{x} - \frac{1}{2} \left( \wf \pd{2}{\wfc}{x} +\wfc \pd{2}{\wf}{x}  \right)
\end{equation}
according to which, and taking into account the confinement boundary conditions, we can rewrite \REq{etc.m} as
\begin{multline}\label{etc.m.p}
   \tds{}{\mom}{t} = \mofix \; +  \\ 
   \sigma\cdot \left\{ \left[ - \pd{}{}{x} \left( \hpibb \pd{}{\wfc\wf}{x} \right) \right]_{x=+L} 
                     - \left[ - \pd{}{}{x} \left( \hpibb \pd{}{\wfc\wf}{x} \right) \right]_{x=-L}   \right\}
\end{multline}
\REqb{etc.m.p} reveals better than \REq{etc.m}, we believe, the role of {\authorcorr the external forces} by clearly identifying the spatial derivative of the probability density as a potential 
\begin{equation*}
   \hpibb \pd{}{\wfc\wf}{x}
\end{equation*}
whose gradients calculated at the boundaries of the spatial domain {\authorcorr characterise the action of the confining walls}  on the particle. 
Such forces are switched off in Ehrenfest's formulation \mbox{$(\sigma = 0)$}; however, if \mbox{$\sigma=1$} then they impact the macroscopic observable momentum exactly as the force due to the physical potential does.
{\authorcorr A} simple way to push this conflictual situation to stand out with incontestable evidence is to consider the {\authorcorr static} case in which the initial wavefunction [\REq{Seq.pib.ic}] coincides with a pure eigenfunction
\begin{equation}\label{iwf}
   f(x) = \psi_{k}(x)
\end{equation}
This initial condition presupposes the following sequence of simplifications.
Only the coefficient corresponding to the eigenfunction survives \mbox{($c_{r}^{} = \delta_{rk},\,c_{s}^{\ast} = \delta_{sk}$)} and the wavefunction general solution simplifies to
\begin{equation}\label{Seq.pib.gs.efk}
   \wf(x,t) = \exp\left(- i \frac{\epsilon_{k} t}{\hbar}\right) \psi_{k}(x)
\end{equation}
The macroscopic observable position [\REq{mo.x}] becomes time-independent
\begin{equation}\label{mo.x.ss}
   \mop = \int_{-L}^{+L} \psikc x \psik dx
\end{equation}
its time derivative [\REq{td-x.ef.d}] vanishes
\begin{equation}\label{td-x.ef.d.ss}
   \tds{}{\mop}{t} = 0
\end{equation}
and so do the macroscopic observable momentum  [\REq{etc.p}]
\begin{equation}\label{etc.p.ss}
   \mom = 0
\end{equation}
and its time derivative [\REq{td-p.ef.d}]
\begin{equation}\label{td-p.ef.d.ss}
   \tds{}{\mom}{t} = 0
\end{equation}
The attentive reader may notice that \REq{mo.p} reduces to
\begin{equation}\label{mo.p.ss}
   \mom = \int_{-L}^{+L} \psikc \hoi \pd{}{\psik}{x} dx
\end{equation}
and perceive an inconsistency between \REqd{etc.p.ss}{mo.p.ss}; such an inconsistency, however, is only apparent and is quickly removed by \REqc{p.iv}{(c)} particularised to the case $s=r$ because it enforces, for any arbitrary $r$, the (rather curious) orthogonality condition
\begin{equation}\label{p.iv.c.s=r}
   \int_{-L}^{+L} \psi_{r}^{\ast} \hoi \pd{}{\psir}{x} dx = 0
\end{equation}
that,  with $r=k$, realigns \REq{mo.p.ss} to \REq{etc.p.ss}.
Now, what happens to \REq{etc.m}? The left-hand side vanishes according to \REq{td-p.ef.d.ss}; the right-hand side follows from \REqc{td-p.ef}{(c)} with $r=s=k$
\begin{equation}\label{etc.m.p.ss}
   0 = \mofixef{k}{k} -
       \sigma \cdot \hpib \left[ \left( \pd{}{\psik}{x} \pd{}{\psikc}{x} \right)_{x=+L} - \left( \pd{}{\psik}{x} \pd{}{\psikc}{x} \right)_{x=-L}\right]
\end{equation}
in which we have introduced the switch $\sigma$ for consistency with \REq{etc.m}.
\REqb{etc.m.p.ss} describes {\authorcorr the} equilibrium between {\authorcorr the force} due to the physical potential and the {\authorcorr external forces; a}s a consequence of such equilibrium, we macroscopically observe the particle standing at the position given by \REq{mo.x.ss}.
\REqcb{tiSeq.pib.rs.iii}{(c)} with \mbox{$s=r=k$} helps us to go a bit further than \REq{etc.m.p.ss} by providing the integral 
\begin{equation}\label{tiSeq.pib.rs.iii.ss}
   \mofixef{k}{k} = \hpib \left[ \left( \pd{}{\psik}{x} \pd{}{\psikc}{x} \right)_{x=+L} - \left( \pd{}{\psik}{x} \pd{}{\psikc}{x} \right)_{x=-L}\right]
\end{equation}
so that \REq{etc.m.p.ss} reduces to the final form
\begin{equation}\label{etc.m.p.ss.f}
   0 = ( 1 - \sigma) \cdot \hpib \left[ \left( \pd{}{\psik}{x} \pd{}{\psikc}{x} \right)_{x=+L} - \left( \pd{}{\psik}{x} \pd{}{\psikc}{x} \right)_{x=-L}\right]
\end{equation}
With a look at \REq{etc.m.p.ss.f}, we conclude that everything is consistent if \mbox{$\sigma=1$}; on the other hand, a conflict becomes evident if \mbox{$\sigma=0$} \textit{and} the {\authorcorr external forces} do not {\authorcorr either} vanish {\authorcorr or compensate}.
In this regard, a numerical test case with a symmetrical potential $V(x)$ will not serve our purpose because the symmetry will {\authorcorr flatten out to zero the right-hand side of} \REq{etc.m.p.ss.f}. 
To convince ourselves, it is sufficient to open Pauling and Wilson's textbook \cite{pw1935} and look at the first four eigenfunctions produced by the uniform potential [Fig.~14-2 (left)] or by the (unconfined) harmonic-oscillator potential [Figs.~11-2 (left) and 11-3 (top)]. 
In the former case, {\authorcorr the external forces compensate because} the squares of the slopes coincide
\begin{subequations} \label{pw}
	\begin{equation}\label{pw.pib}
	   \left( \pd{}{\psik}{x} \pd{}{\psikc}{x} \right)_{x=+L} = \left( \pd{}{\psik}{x} \pd{}{\psikc}{x} \right)_{x=-L}
	\end{equation}
while in the latter case {\authorcorr the external forces vanish because} the slopes vanish
	\begin{equation}\label{pw.uho}
	   \left( \pd{}{\psik}{x} \right)_{x=\pm\infty} = 0
	\end{equation}
\end{subequations}
Thus, a test case with unsymmetrical potential is advisable and, therefore, we have considered an electric charge $q$ in a uniform electrostatic field $E$ with potential
\begin{equation}\label{pot.esf}
   V(x) = - qEx
\end{equation} 
This potential satisfies \REq{pot.eq} and is safely beyond the reach of the criticisms discussed in \Rse{intro} in connection with that condition.
The macroscopic observable
\begin{equation}\label{mo.V}
   \mo{V} = \mopi = - q E \mop = V\left( \mop \right)
\end{equation}
comes to coincide with the potential as a function of the macroscopic observable position; consequently, and taking in due account the wavefunction normalisation [\REq{wfn}], the {\authorcorr macroscopically observable} force can be expressed as follows
\begin{equation}\label{mo.F.cef}
   \mofix = - \pd{}{\mo{V}}{\mop} = qE
\end{equation}
and \REqq{et.c} become
\begin{subequations} \label{et.w}
	\begin{align}
	  \tds{}{\mop}{t} & = \frac{1}{m}\mom  \label{etw.p} \\[.5\baselineskip]
	  \tds{}{\mom}{t} & = - \pd{}{\mo{V}}{\mop} 
	  \textcolor{grey}{\; - \;  \sigma\cdot\hpib \left[ \left( \pd{}{\wf}{x} \pd{}{\wfc}{x} \right)_{x=+L} - \left( \pd{}{\wf}{x} \pd{}{\wfc}{x} \right)_{x=-L}\right]}    \label{etw.m}
	\end{align}
\end{subequations}
We have dimmed the {\authorcorr external forces} in \REq{etw.m} because they are invisible in Ehrenfest's formulation [\REqq{pet}] and to bring forth the evidence that, with the dimmed terms unseen, not only the macroscopically observable {\authorcorr statics/dynamics} of the electric charge would {\authorcorr comply with} the classical-mechanics prescript but also, and more importantly, that quantum mechanics would have really nothing to say in that respect because \REqq{et.w} would be in a mathematically closed form that could be integrated in time without any connection whatsoever with the underlying microscopically quantic substrate.
Yet, the dimmed terms are undeniably there and, through them, quantum mechanics has its saying{\authorcorr; w}e will find out in next section if and how. 
Well, it looks like we are settling for a rather attractive test case.

\subsubsection{Electric charge in uniform electrostatic field.\label{ecef}}

With the electrostatic potential of \REq{pot.esf}, the time-independent Schr\"{o}dinger equation [\REq{tiSeq.pib}] takes the explicit form
\begin{subequations}\label{tiSeq.pib.cef}
	\begin{equation}\label{tiSeq.pib.cef.de}
	   - \hpib\pd{2}{\psi}{x} -qEx\psi = \epsilon \psi
	\end{equation}
that we must integrate with the confinement boundary conditions
	\begin{equation}\label{tiSeq.pib.cef.bc}
       \psi(-L) = \psi(+L)=0
	\end{equation}
and the normalisation condition [\REq{efn}]
	\begin{equation}\label{tiSeq.pib.cef.n}
       \int_{-L}^{+L} \!\!\!\psi^{\ast}\psi \, dx = 1
	\end{equation}
\end{subequations}
We rephrase the eigenvalue problem by referring the $x$ coordinate to the semi-length of the one-dimensional box
\begin{subequations}\label{ep.nd}
	\begin{equation}\label{x.nd}
	   x = \xi\cdot L 
	\end{equation}
	and by scaling the eigenfunctions
	\begin{equation}\label{ef.nd}
	    \psi \rightarrow \frac{\psi}{\sqrt{2L}}
	\end{equation}
\end{subequations}
In this way, \REqq{tiSeq.pib.cef} turn into the nondimensional form{\authorcorr s}
\begin{align}\label{tiSeq.pib.cef.nd}
   - \pd{2}{\psi}{\xi} - \alpha \xi \psi = \beta \psi  &&   \psi(-1) = \psi(+1) = 0  &&   \frac{1}{2} \int_{-1}^{+1} \!\!\!\psi^{\ast}\psi \, d\xi = 1
\end{align}
which include two characteristic numbers 
\begin{subequations}\label{cn}
	\begin{align}
	   \alpha = & \frac{2mqEL^{3}}{\hbar^{2}}         \label{cn.a}\\[2ex]
	   \beta  = & \frac{2mL^{2}}{\hbar^{2}} \epsilon  \label{cn.b}
	\end{align}
\end{subequations}
It is rather straightforward to prove from \REqq{tiSeq.pib.cef.nd} that
\begin{equation}\label{nalpha}
   \psi(\xi,-\alpha) = \psi(-\xi,\alpha)
\end{equation}
and, therefore, we concentrate only on the solutions with positive $\alpha$; they describe the situations in which electric charge and electric field have same sign.
For the {\authorcorr numerical} solution of our mathematical problem, we have utilised a method based on high-order finite differences \cite{pa2011jnaiam,pa2015cnsns,pa2020lncs} implemented in the code HOFiD\_MSP that can solve multiparameter spectral BV-ODE problems; in our calculations we have always utilised 8th-order formulae on a grid composed of 1001 equally spaced points on the interval $[-1,\footplus 1]$.

As it is well known, if the electric field is absent ($\alpha=0$) then \REqq{tiSeq.pib.cef.nd} produce analytical eigenfunctions
\begin{equation}\label{aef}
   \psi_{k}(\xi) = \sqrt{2} \sin\left[  \frac{k\pi}{2} \left( \xi + 1 \right) \right]
\end{equation}
and eigenvalues
\begin{equation}\label{aev}
   \beta_{k} = \left( \frac{k\pi}{2} \right)^{2}
\end{equation}
With \REq{aev} in hand, the dimensional energy levels
\begin{equation}\label{ael}
   \epsilon_{k} = \left( \frac{k\pi}{2} \right)^{2} \frac{\hbar^{2}}{2mL^{2}} = \frac{k^{2}h^{2}}{8m(2L)^{2}}
\end{equation}
are easily retrieved from \REq{cn.b}; to be on the safe side, we have verified \REq{ael} against \mbox{Eq.~(14-7)} at page 97 of Pauling and Wilson's textbook \cite{pw1935} (notation correspondence: $k \rightarrow n_{x}$; $2L\rightarrow a$).
We have used this analytical solution [\REq{aef}] to validate our numerical method; the outcome of the validation exercise, consisting in the superposition of analytical and numerical calculations for the first four eigenfunctions ($k=1,2,3,4$), is illustrated in \Rfi{vas} and is very satisfactory.
\newlength{\ftwidth}  \setlength{\ftwidth}{\textwidth}   \addtolength{\ftwidth}{-\mathindent}

\newlength{\sfsize}  \setlength{\sfsize}{.49\ftwidth}

\begin{figure}[h]
   \hspace*{\mathindent}
   \resizebox{\sfsize}{!}{\includegraphics*[trim=10 25 60 60]{\figdir/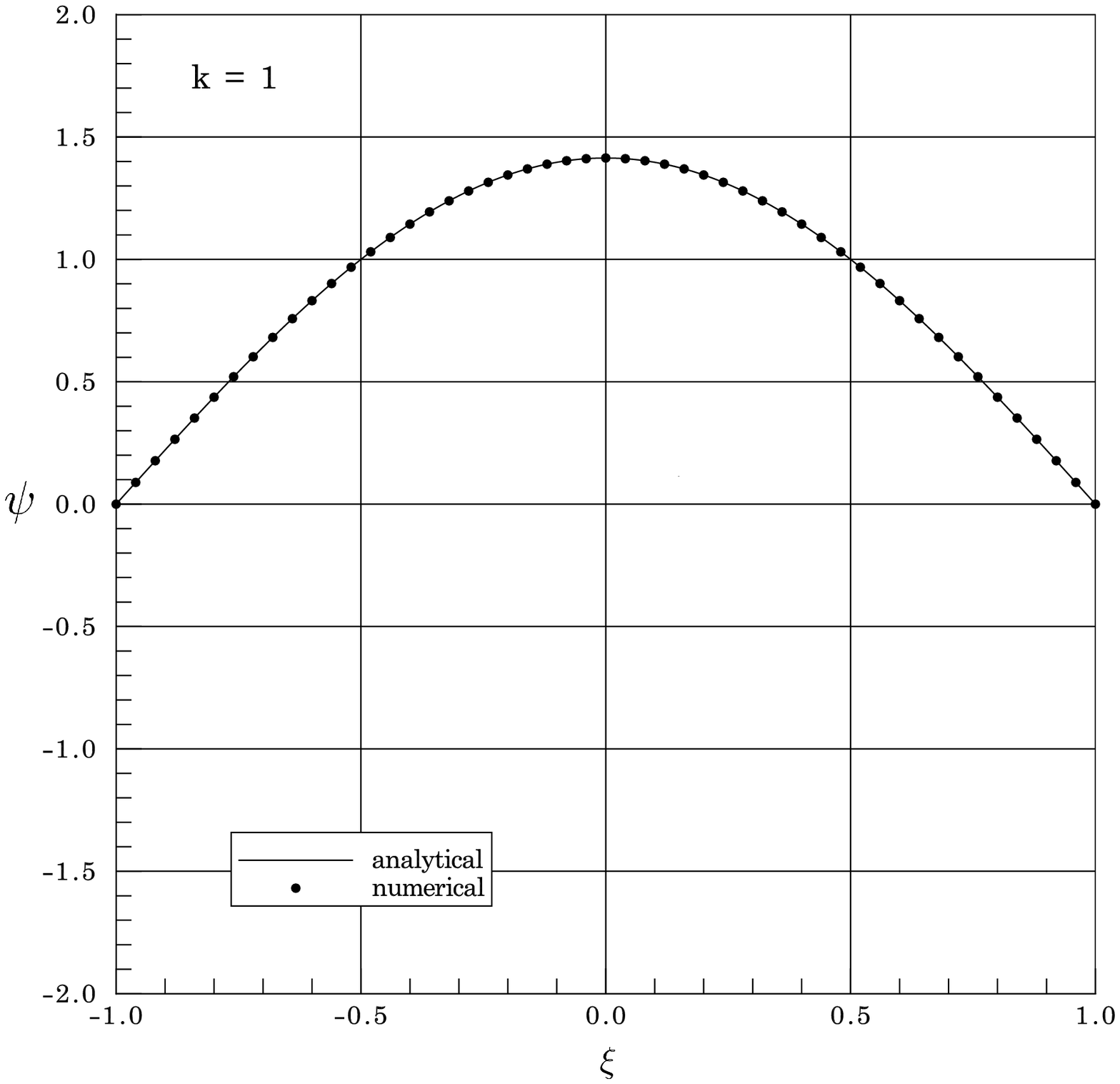}} 
   \resizebox{\sfsize}{!}{\includegraphics*[trim=10 25 60 60]{\figdir/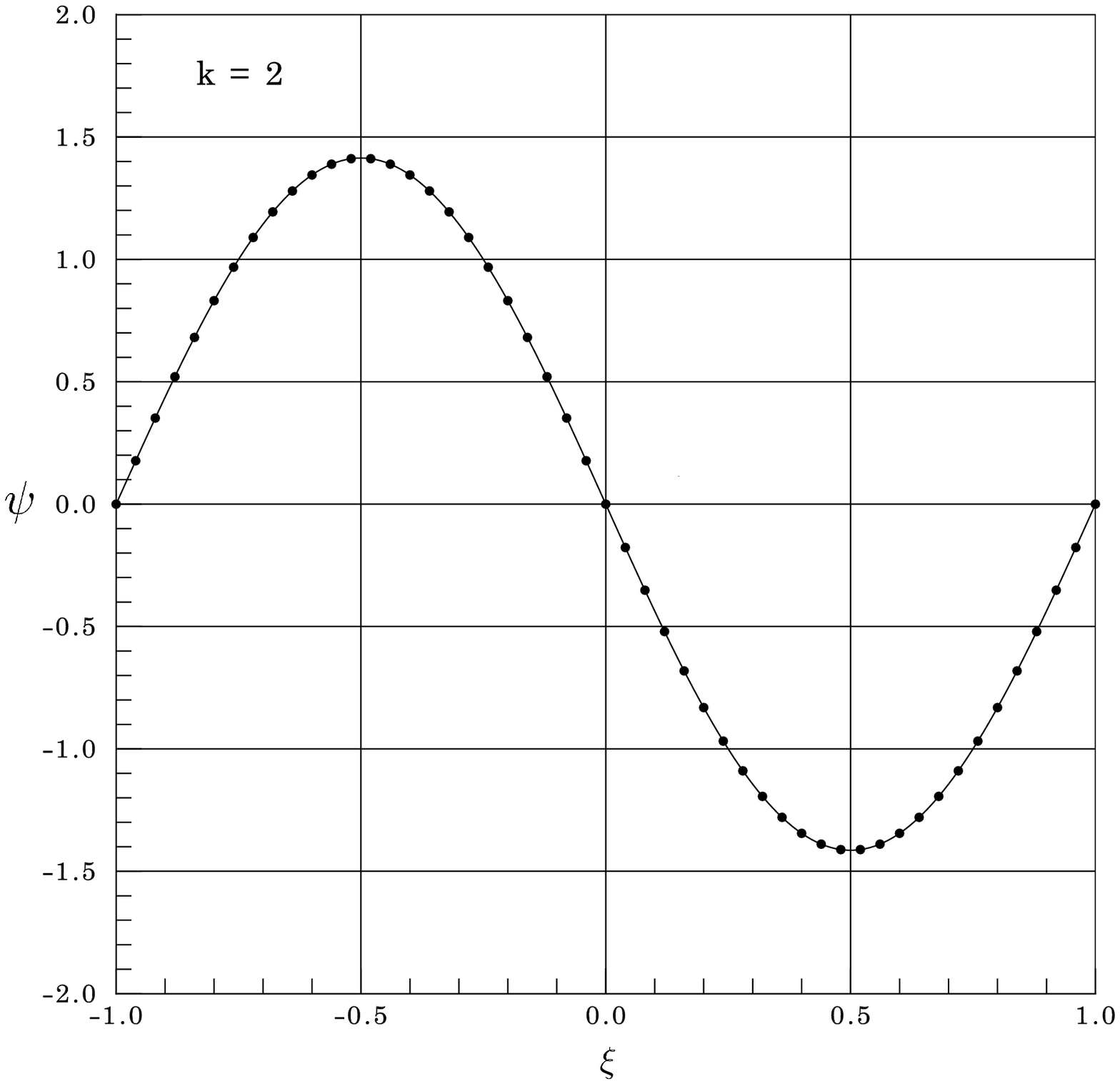}} \\ 
   \hspace*{\mathindent}
   \resizebox{\sfsize}{!}{\includegraphics*[trim=10 25 60 60]{\figdir/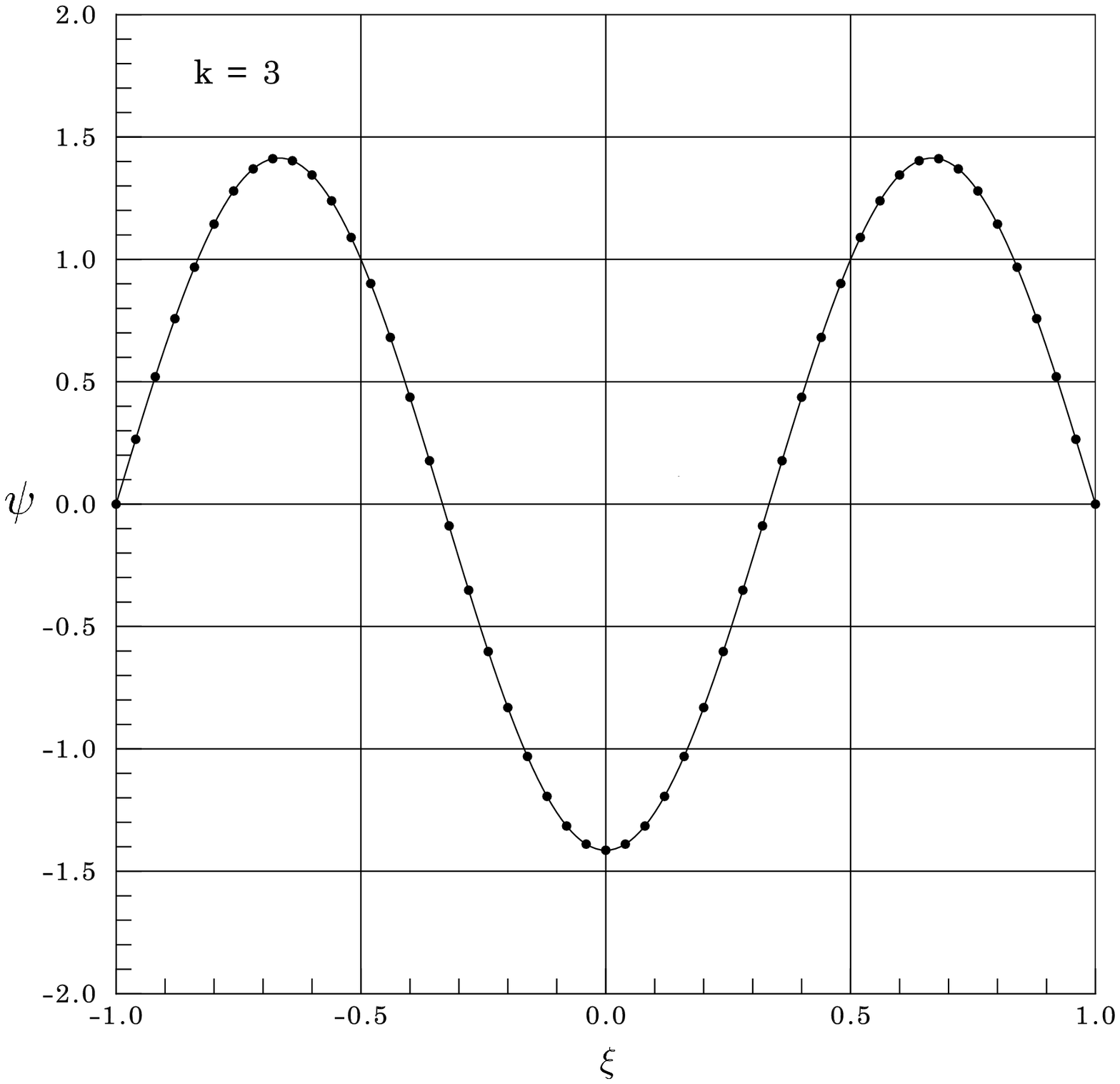}} 
   \resizebox{\sfsize}{!}{\includegraphics*[trim=10 25 60 60]{\figdir/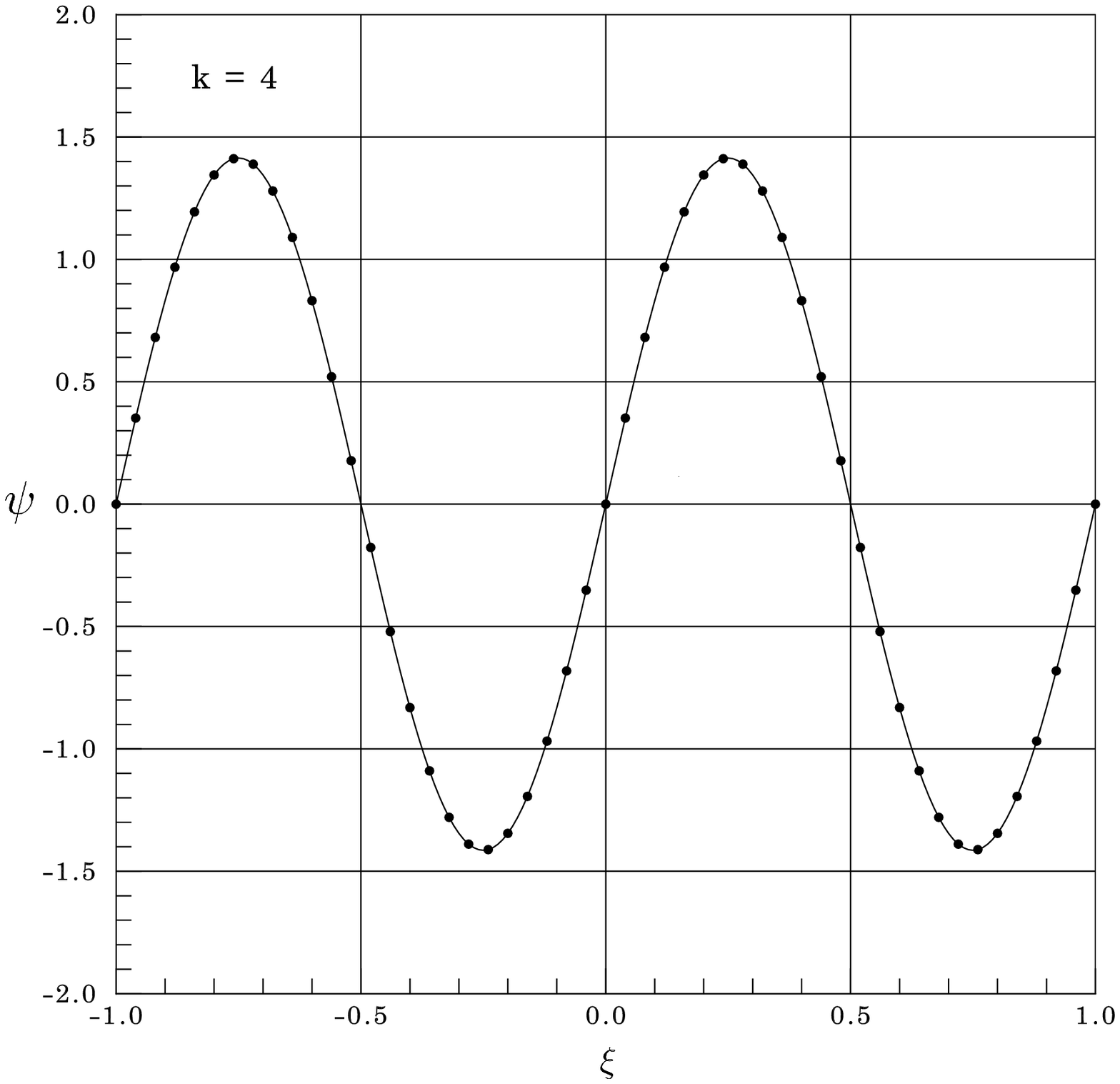}}  
   \caption{Validation exercise with the first four eigenfunctions of the analytical solution [\REq{aef}] corresponding to $\alpha=0$.\label{vas}}
\end{figure}

With the validation test positively passed, we have carried out calculations for the first four stationary states ($k=1,2,3,4$) with the electric field switched on at \mbox{$\alpha=10,100$}.
The eigenvalues are listed in \Rta{table} as direct output from the calculation {\authorcorr (2nd column)} and scaled with the numerical factor $4/\pi^{2}$ {\authorcorr (3rd column)} in analogy to \REq{aev}.%
\begin{table}[h]
   \hspace*{\mathindent}
   \resizebox{\ftwidth}{!}{\includegraphics*{\figdir/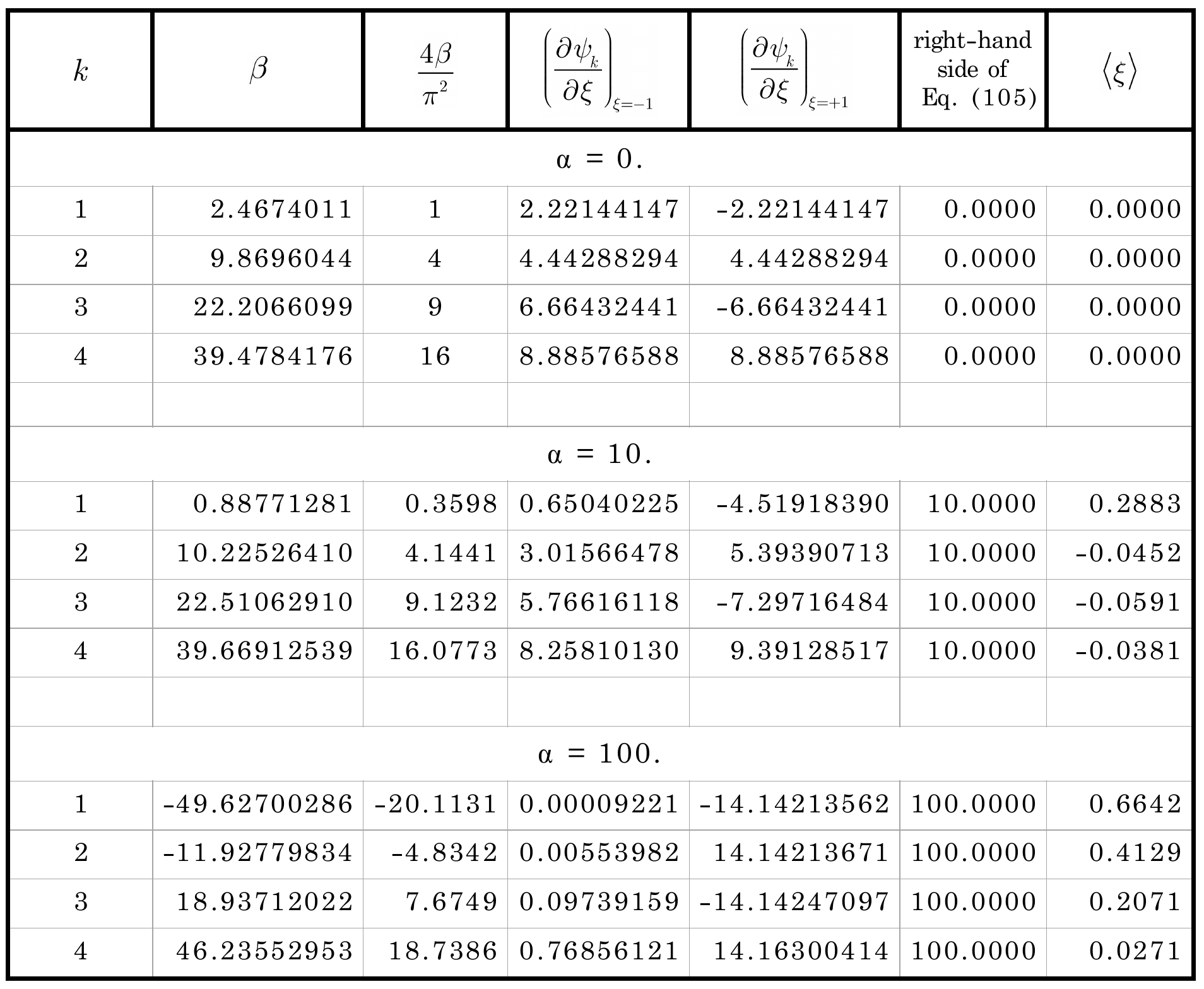}}  
   \caption{Eigenvalues, eigenfunctions' derivatives at the boundaries, and macroscopic observable position for the first four stationary states.\label{table}}
\end{table}
Eigenfunctions and probability densities are shown, respectively, in \Rfid{ef1-4.cef}{pd1-4.cef}. 
The graphs give ample evidence of how the electric field breaks the symmetries, induces a general rightward ($\alpha>0$) shift, and makes uneven maxima and minima.
Also, its impact on the eigenfunctions' derivatives at the boundaries, which are of particular importance to us, is evident.
\begin{figure}[h]
   \hspace*{\mathindent}
   \resizebox{\sfsize}{!}{\includegraphics*[trim=10 25 60 60]{\figdir/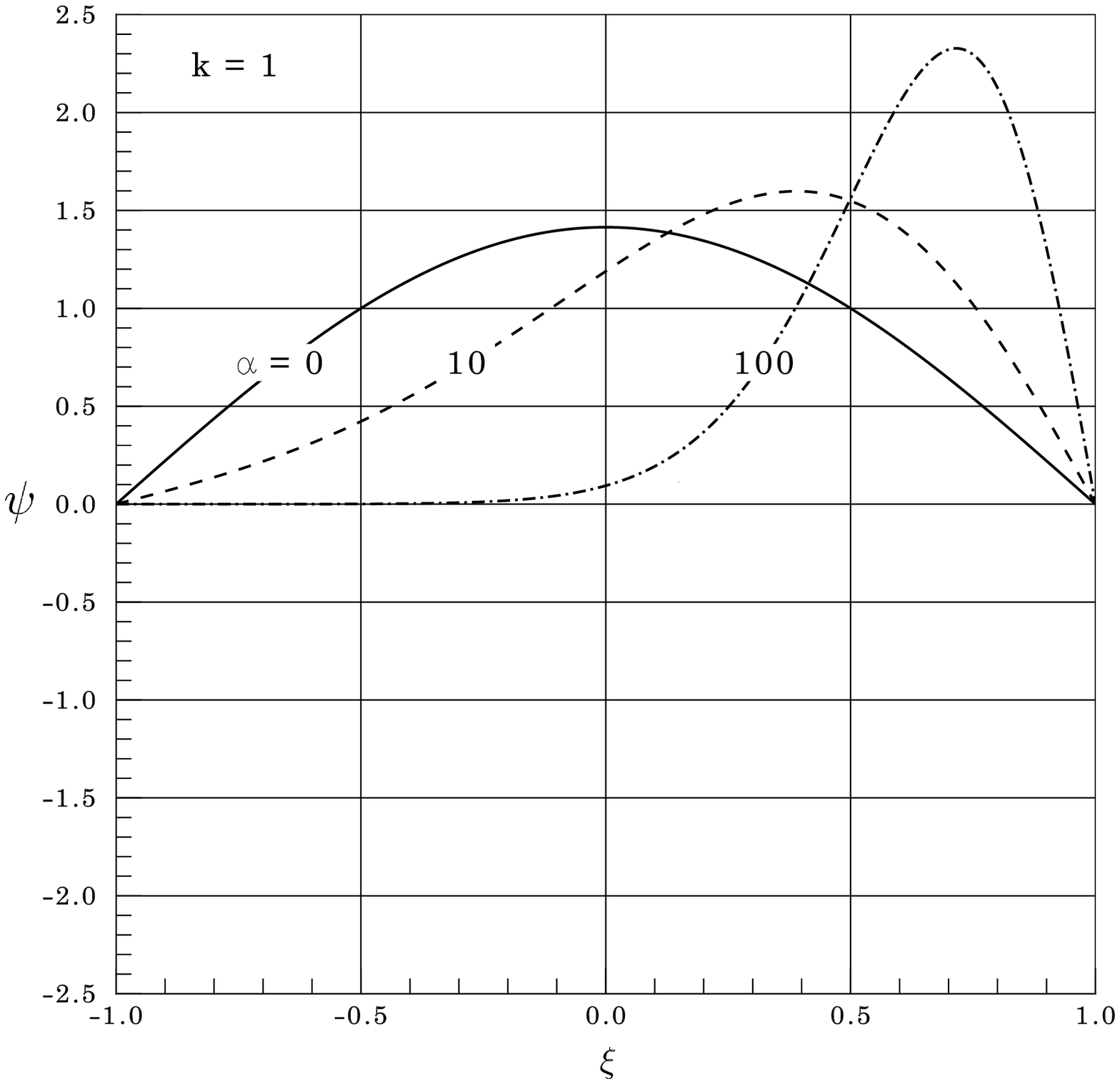}} 
   \resizebox{\sfsize}{!}{\includegraphics*[trim=10 25 60 60]{\figdir/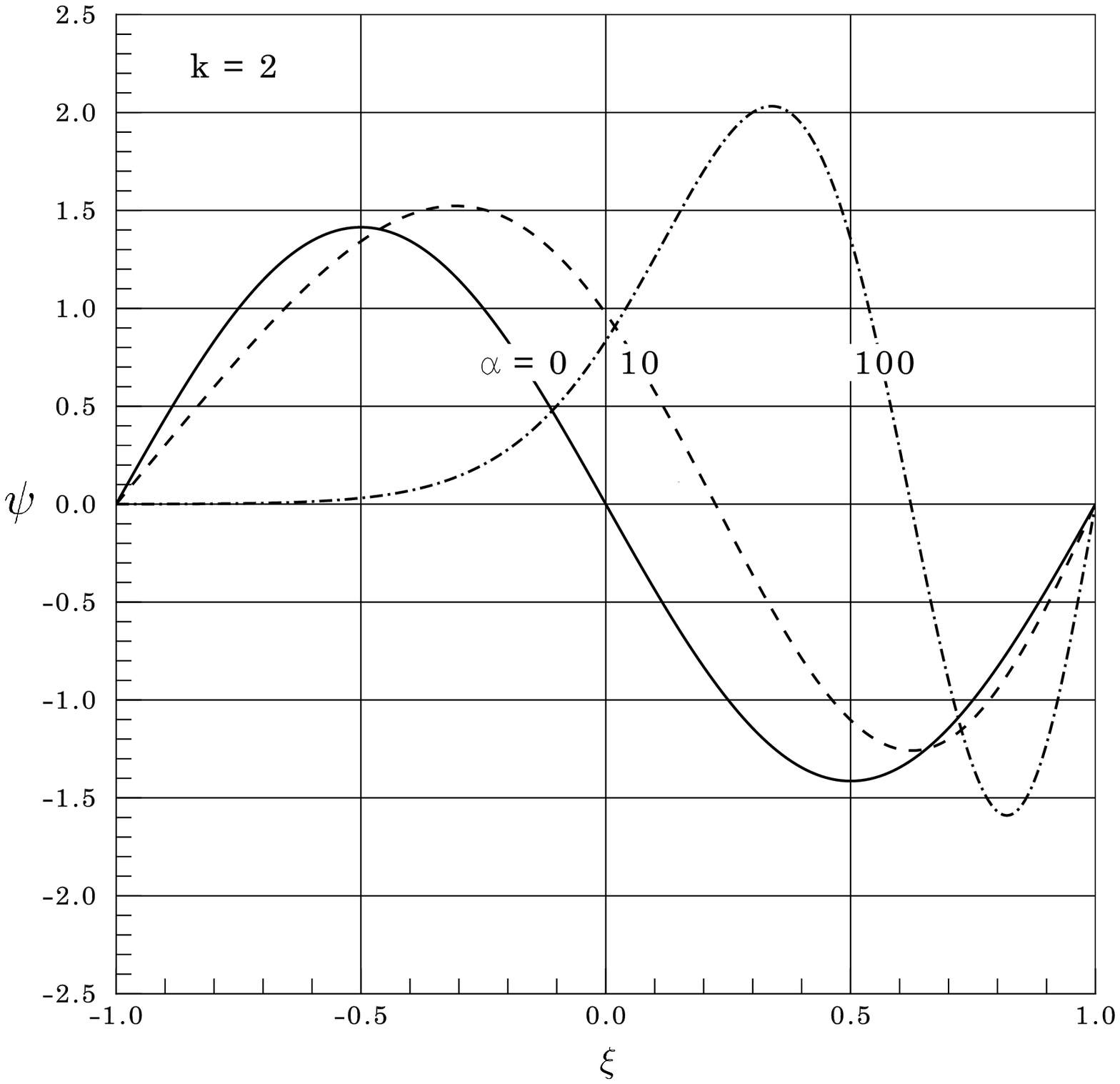}} \\ 
   \hspace*{\mathindent}
   \resizebox{\sfsize}{!}{\includegraphics*[trim=10 25 60 60]{\figdir/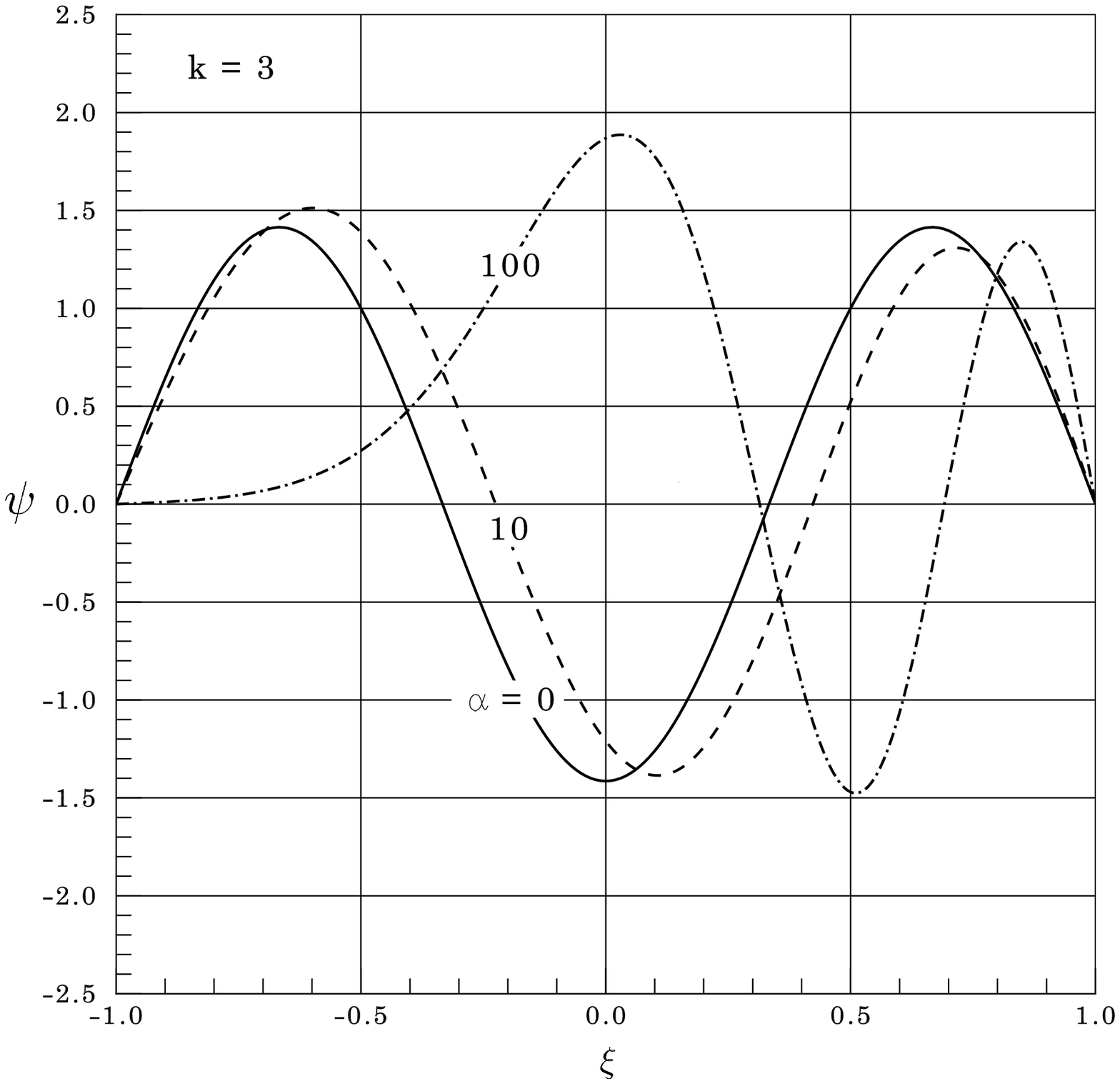}} 
   \resizebox{\sfsize}{!}{\includegraphics*[trim=10 25 60 60]{\figdir/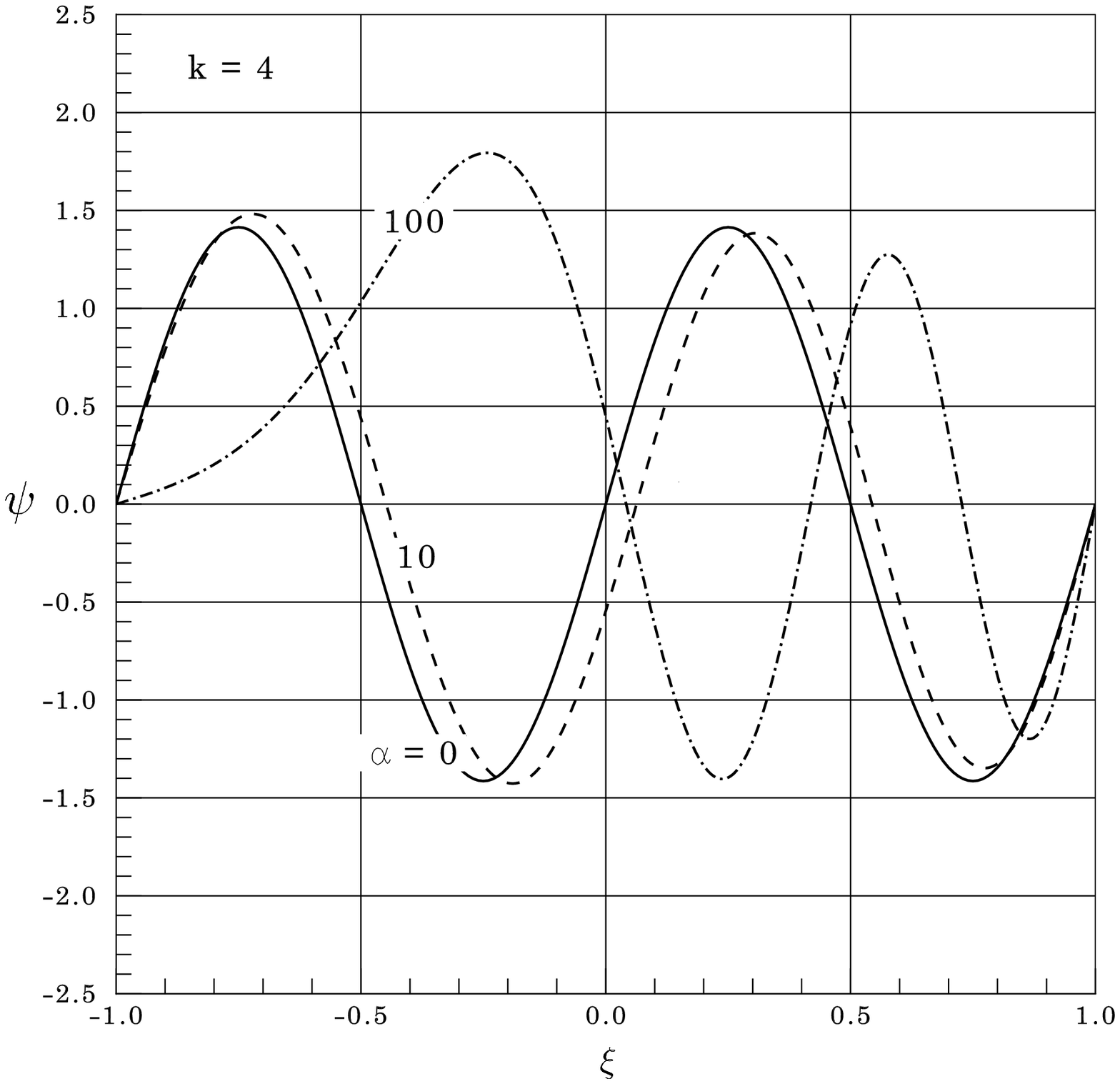}}     
   \caption{The first four eigenfunctions of an electric charge in uniform electrostatic field.\label{ef1-4.cef}}
\end{figure}
\begin{figure}[h]
   \hspace*{\mathindent}
   \resizebox{\sfsize}{!}{\includegraphics*[trim=10 25 60 60]{\figdir/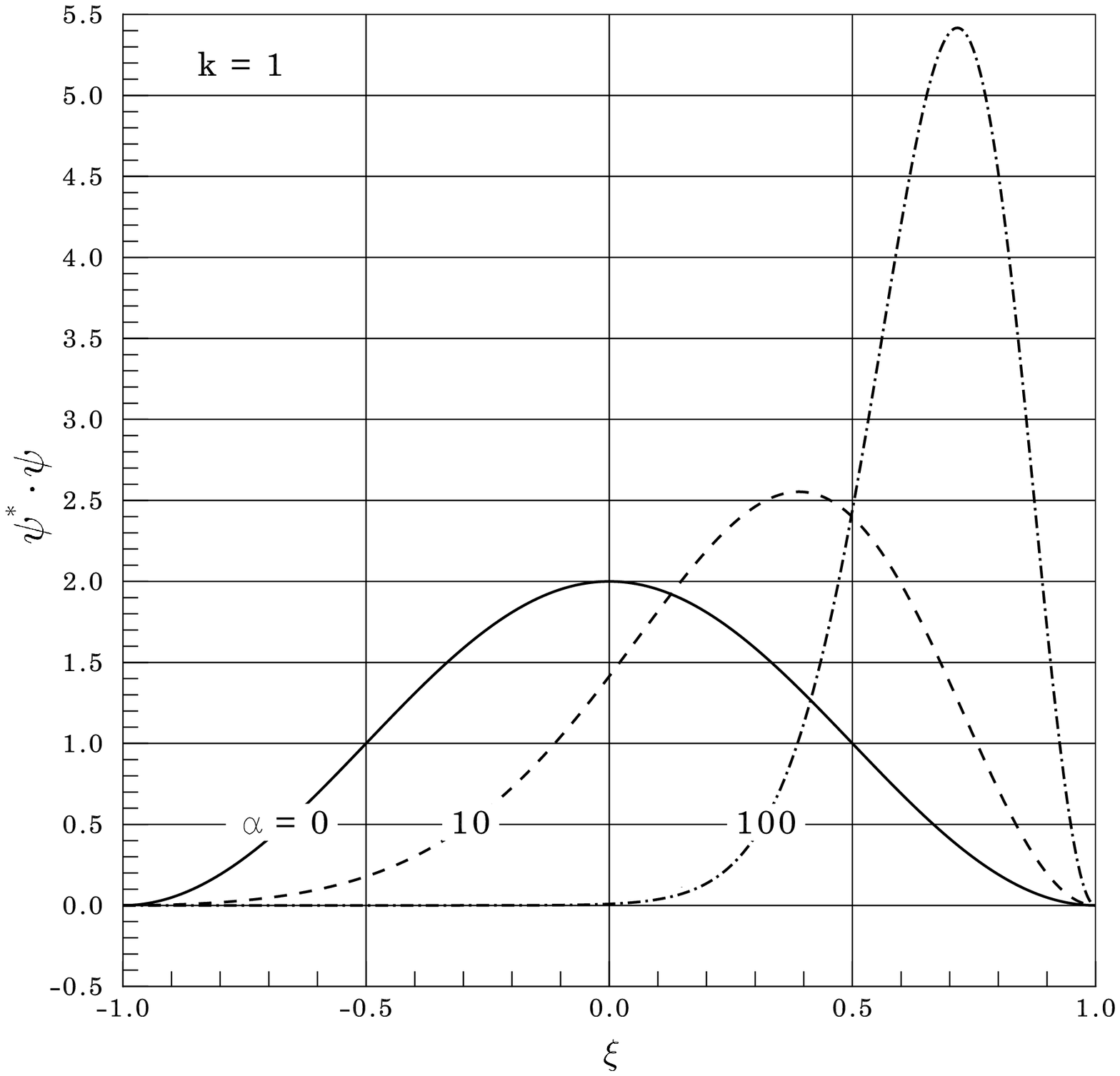}} 
   \resizebox{\sfsize}{!}{\includegraphics*[trim=10 25 60 60]{\figdir/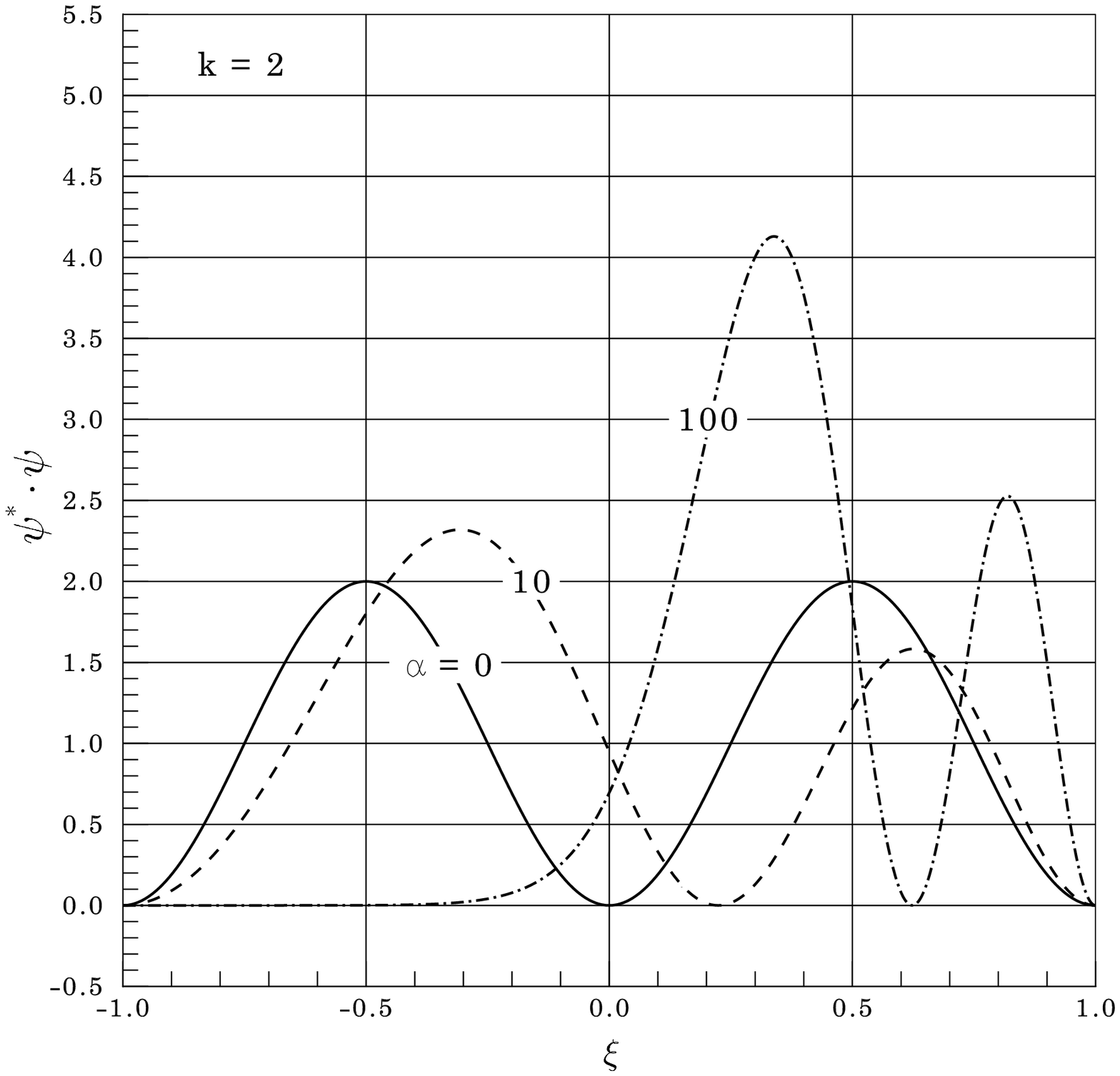}} \\ 
   \hspace*{\mathindent}
   \resizebox{\sfsize}{!}{\includegraphics*[trim=10 25 60 60]{\figdir/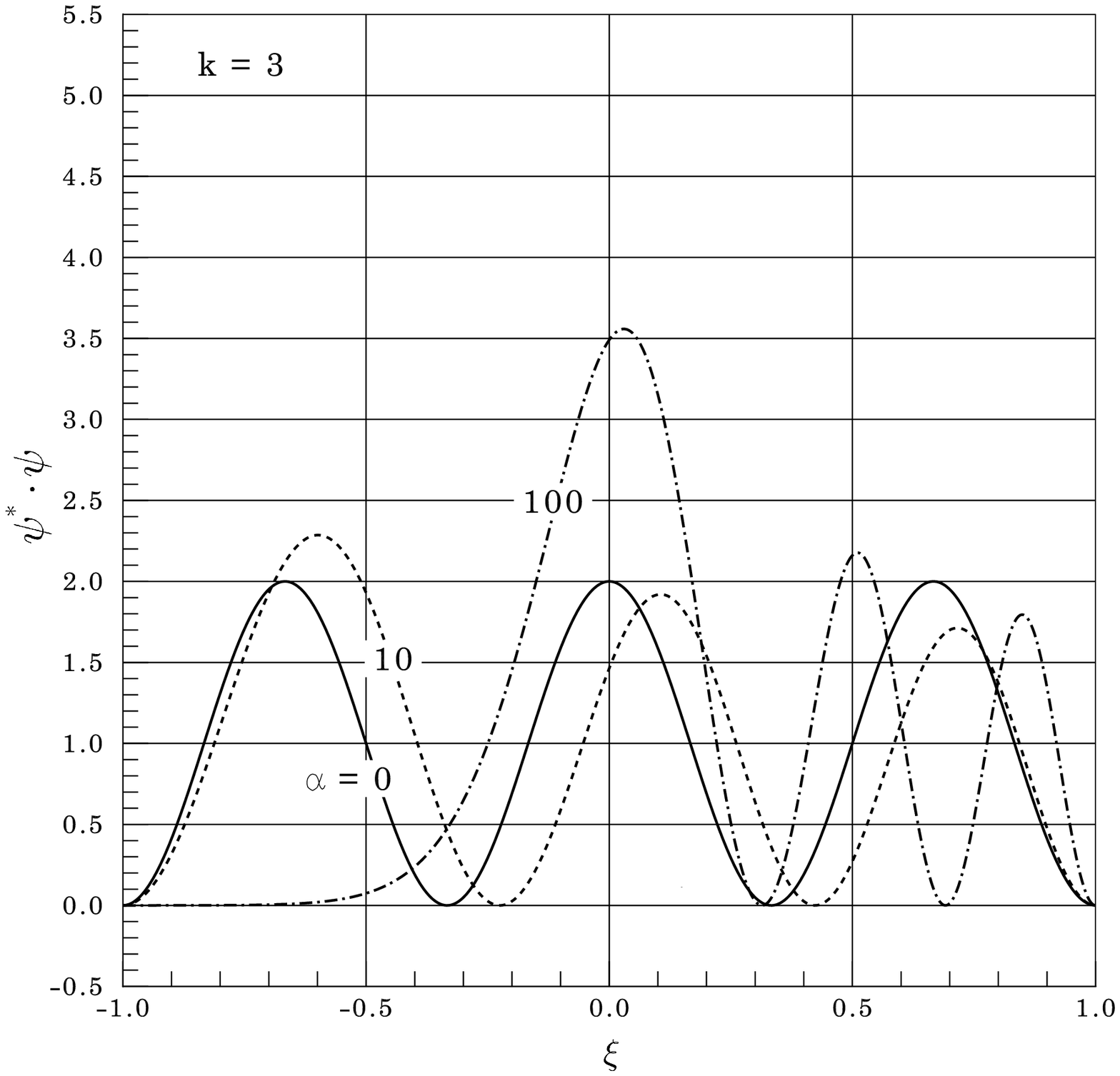}} 
   \resizebox{\sfsize}{!}{\includegraphics*[trim=10 25 60 60]{\figdir/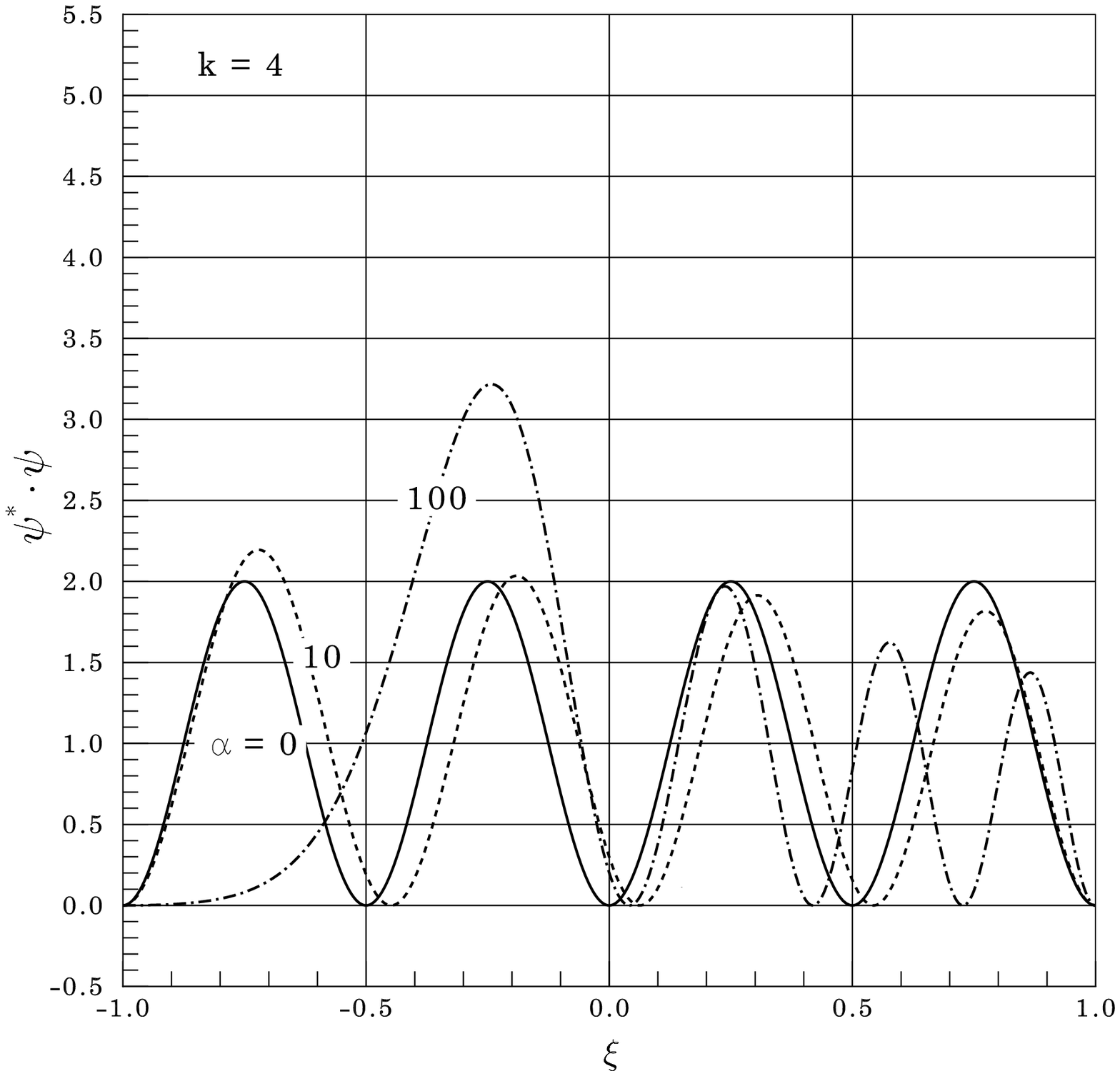}}     
   \caption{Probability density {\authorcorr for} the first four stationary states of an electric charge in uniform electrostatic field.\label{pd1-4.cef}}
\end{figure}
Let us rewrite here \REq{etc.m.p.ss.f} in nondimensional form [\REqq{ep.nd}]
\begin{equation}\label{etc.m.p.ss.f.nd}
   0 = ( 1 - \sigma) \cdot \left[ \left( \pd{}{\psik}{\xi} \pd{}{\psikc}{\xi} \right)_{\xi=+1} - \left( \pd{}{\psik}{\xi} \pd{}{\psikc}{\xi} \right)_{\xi=-1}\right]
\end{equation}
When $\alpha=0$ we can take the derivative of \REq{aef} 
\begin{equation}\label{daef}
   \pd{}{\psi_{k}}{\xi} = \frac{k\pi}{\sqrt{2}} \cos\left[  \frac{k\pi}{2} \left( \xi + 1 \right) \right]
\end{equation}
calculate it at the boundaries
\begin{align}\label{daefb}
   \left( \pd{}{\psi_{k}}{\xi} \right)_{\xi=-1} = \frac{k\pi}{\sqrt{2}} &&  \left( \pd{}{\psi_{k}}{\xi} \right)_{\xi=+1} = \frac{k\pi}{\sqrt{2}} \cos(k\pi) 
\end{align}
and obtain the analytical verification of \REq{pw.pib} that we had anticipated on the basis of graphical arguments deduced from \mbox{Fig. 14-2 (left)} in Pauling and Wilson's textbook \cite{pw1935}, also reproduced in our \Rfi{vas}.
Thus, \REq{etc.m.p.ss.f.nd} is satisfied regardless of which value the switch $\sigma$ is set to.
{\authorcorr
With due account of \REqq{ep.nd} and \REq{ael}, \REqq{daefb} allow to retrieve easily the dimensional expression
\begin{equation}\label{ef.dv}
   \hpib \left( \pd{}{\psik}{x} \pd{}{\psikc}{x} \right)_{x=\pm L} \rightarrow \hpib \frac{1}{2L^{3}}\left( \pd{}{\psik}{\xi} \pd{}{\psikc}{\xi} \right)_{\xi=\pm 1}
   = 2\,\frac{\epsilon_{k}}{(2L)} 
\end{equation}
of the external forces, in agreement with the result obtained by De Vincenzo \cite{sdv2013jop,sdv2013rbef} with a similar calculation. 
The same expression, as duly mentioned by De Vincenzo, was also proposed by ter Haar \cite{dth1964} (problem 25 at page 5; solution at page 88) although he followed a completely different path.
{\refereeone Ter Haar pointed out that and, above all, clearly explained how the expression given in \REq{ef.dv} [corresponding to his \mbox{Eq.~(10)} at page 91] ``retains its form exactly in classical mechanics''.}}

When \mbox{$\alpha\neq 0$}, we have calculated the derivatives numerically; they are listed in \Rta{table}.
A quick numerical check with the values collected therein shows that the {\authorcorr external-force difference} in \REq{etc.m.p.ss.f.nd} differ{\authorcorr s} from zero and, therefore, setting the switch to \mbox{$\sigma=0$} (Ehrenfest's mode) brings up an incontrovertible conflict.
An additional check of confidence that \mbox{$\sigma=1$} is the correct choice is offered by \REq{tiSeq.pib.rs.iii.ss} after we substitute in it the electrostatic potential and then turn to the  nondimensional form [\REqq{ep.nd} and \REq{cn.a}]
\begin{equation}\label{tiSeq.pib.rs.iii.nd}
   \alpha = \frac{1}{2} \left[ \left( \pd{}{\psik}{\xi} \pd{}{\psikc}{\xi} \right)_{\xi=+1} - \left( \pd{}{\psik}{\xi} \pd{}{\psikc}{\xi} \right)_{\xi=-1}\right]
\end{equation}
\REqb{tiSeq.pib.rs.iii.nd} is {\authorcorr the nondimensional} way to say that the electrical force is balanced by the {\authorcorr external} forces.
The right-hand side's values produced by the numerically obtained derivatives listed in \Rta{table} are contained in the second column from right in the same table. 
The match with {\authorcorr the values of} $\alpha$ is indisputable; as a matter of fact, we were really surprised to see how well those apparently patternless numerically determined values of the {\authorcorr external forces} comply with the equilibrium dictated by \REq{tiSeq.pib.rs.iii.nd}.
The verdict ruled by the non-vanishing of the {\authorcorr external forces} in \REq{etc.m.p.ss.f.nd} and by their compliance with \REq{tiSeq.pib.rs.iii.nd} is that the correct position of the switch [\REq{ns}] is \mbox{$\sigma=1$}; this is the {\authorcorr numerical} corroboration of the theory we were looking for.

As a consequence of the equilibrium between the electric force and the {\authorcorr external} forces, the electric charge does not move and, therefore, does not radiate.
It stands quietly at the {\authorcorr nondimensional} macroscopically observable position {\authorcorr [\REq{mo.x.ss}]}
\begin{equation}\label{mo.x.ss.nd}
   \mopnd = \frac{1}{2}\int_{-1}^{+1} \!\!\psikc\,\xi\,\psik\,d\xi
\end{equation}
The values produced by \REq{mo.x.ss.nd} for the first four stationary states are collected in the rightmost column of \Rta{table}.
Obviously, the electric-charge equilibrium position always coincides with the center of the box ($\mopnd = 0$) if the electric field is absent ($\alpha=0$); the analytical confirmation is achieved by substituting \REq{aef} in \REq{mo.x.ss.nd} and executing the integral.
In the case \mbox{$\alpha=10$}, the position shifts expectedly to the right ($\mopnd > 0$) for the ground state \mbox{($k=1$)} but surprisingly to the left ($\mopnd < 0$) for the excited states \mbox{($k=2,3,4$)}.
In the case \mbox{$\alpha=100$}, the {\authorcorr values} indicate a clear rightward shift for all the considered stationary states.
In order to achieve a global and better insight about the influence of the electric field, we have scanned numerically the output of \REq{mo.x.ss.nd} with $\alpha$ ranging in the interval $[0,100]$ and plotted the results in \Rfi{nmop}. 
\begin{figure}[h]
   \hspace*{\mathindent}
   \resizebox{\ftwidth}{!}{\includegraphics*[trim=10 25 60 60]{\figdir/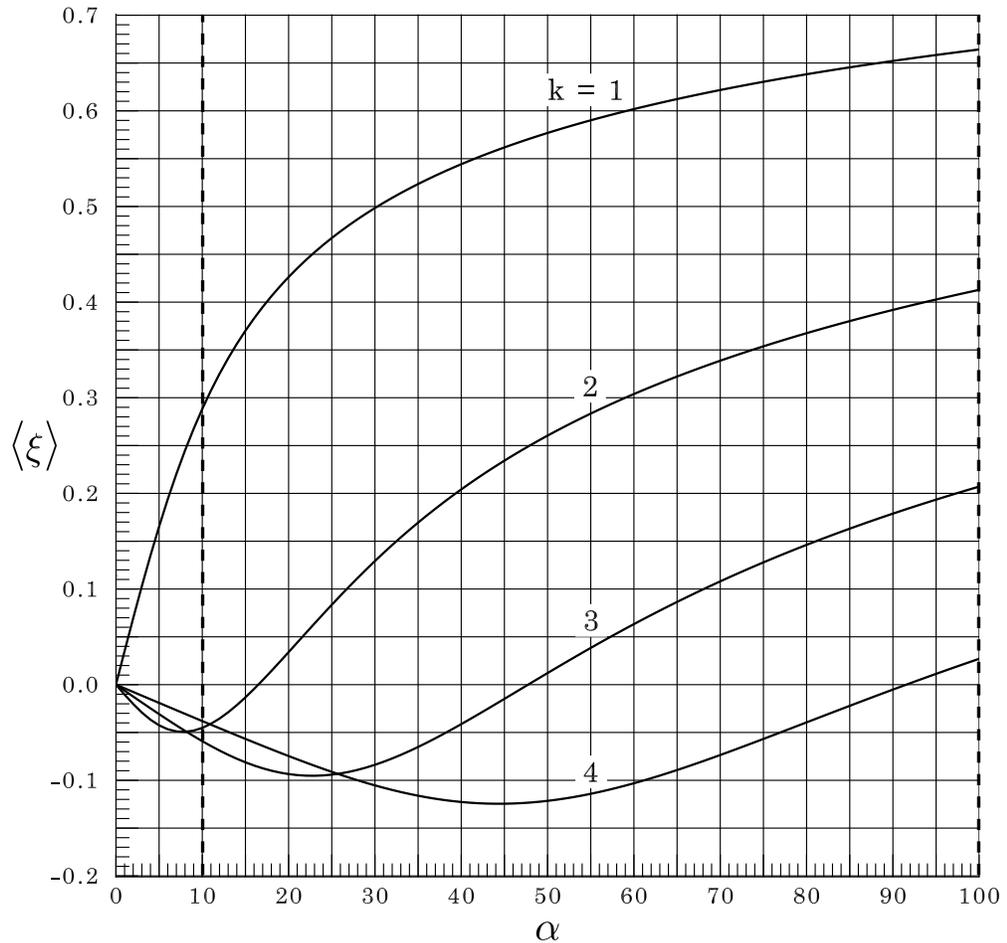}} 
   \caption{Nondimensional macroscopic observable position versus electric field.\label{nmop}}
\end{figure}
They confirm that the position shifts monotonically rightward for the ground state but they {\authorcorr also give evidence of} the existence of a leftward-position range (between the minima and the zero level in \Rfi{nmop}) for the excited states.
This may appear surprising at first sight because it would look like as if, for example, a positive electric charge be repelled by an electric field pointing in the positive direction of the $x$ axis{\authorcorr. 
B}ut we should not both be mislead by our intuition and overlook the fact that the electric force is not the only one present: the electric-charge position [\REq{mo.x.ss.nd}] is decided by the equilibrium between the electric force and the {\authorcorr external forces} [\REqd{etc.m.p.ss}{tiSeq.pib.rs.iii.nd}].

A similar want of better insight is triggered by the eigenvalues shown in \Rta{table}.
With respect to an increase of electric-field intensity, the {\authorcorr values} reported therein indicate a monotonic decrease for the ground state, the existence of a maximum for the stationary states \mbox{$k=2,3$} and a monotonic increase for the stationary state \mbox{$k=4$}.
The systematic series of calculations with our code HOFiD\_MSP, in eigenvalue-only mode, for $\alpha$ ranging in the interval $[0,100]$ has produced the results shown in \Rfi{nsev}.
\begin{figure}[h]
   \hspace*{\mathindent}
   \resizebox{\ftwidth}{!}{\includegraphics*[trim=10 25 60 60]{\figdir/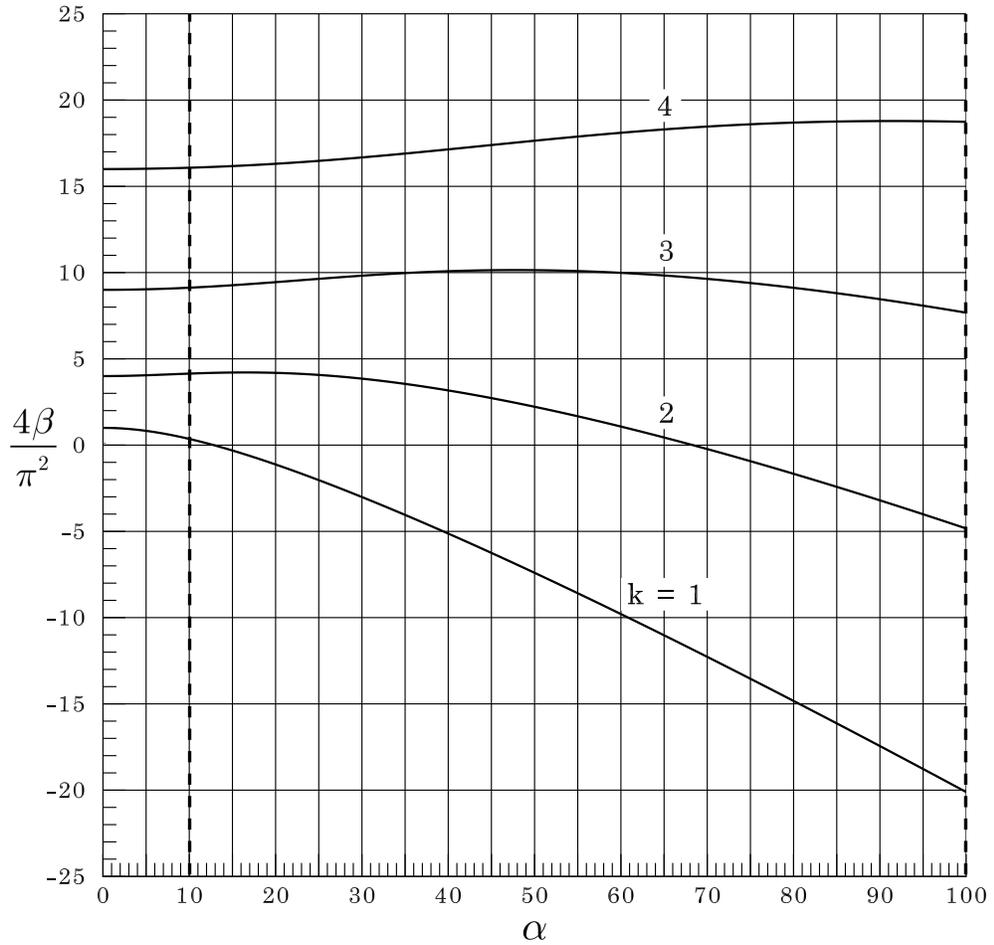}} 
   \caption{Nondimensional scaled eigenvalues versus electric field.\label{nsev}}
\end{figure}
Indeed, each stationary state features a (rather shallow than peaked) maximum eigenvalue. 
The ground-state maximum eigenvalue corresponds to the absence of the electric field \mbox{($\alpha=0$)}; then the eigenvalues decrease monotonically as suggested by \Rta{table}.
The monotonicity of the stationary state \mbox{$k=4$} was only apparent.

\section{Conclusions\label{concl}}

The analysis described in the preceding sections has arisen from the interest to investigate the incompleteness of Ehrenfest's theorem and the necessity to resolve the doubt regarding whether or not the hermiticity of the hamiltonian is sufficient to make the integral $I_{\,\Omega}$ [\REq{Iomega}] vanish unconditionally and, thus, to permit the simplification of a macroscopic-observable time derivative from \REq{motd} to \REq{motd.s}, a notorious praxis in quantum-mechanics textbooks.
We have carried out the study by considering the simple quantum system ``particle in a one-dimensional box''.
We have emphasised the key role of the boundary conditions [\REqq{bc}] to certify a given operator as hermitean and have shown explicitly that, the hermiticity of the hamiltonian notwithstanding, the integral $I_{\,\Omega}$ may not vanish [\REqd{Ix.pib.bc}{Ip.pib.bc}] necessarily; for a given operator, the final word in that respect is left to the boundary conditions, once again.
The immediate consequence is that the claimed generality of the Ehrenfest's formulation [\REqq{pet}] of his famous theorem decays because additional boundary terms [\REqd{td-x}{td-p}] appear. 
In this regard, we have systematically and consistently retrieved the same expressions [\REq{td-x-hill} and \REqc{td-p}{(c)}] of the boundary terms already obtained by Hill \cite{rnh1973ajp}{\authorcorr,} Alonso \etal \cite{va2000inc,va2001pla} {\authorcorr and De Vincenzo \cite{sdv2013jop,sdv2013rbef}}.
{\authorcorr We have investigated and discussed the physical meaning of the boundary terms for the (c) and (p) cases in \Rse{postc}.}
With regard to the time derivative of the macroscopic observable momentum {\authorcorr [\REqc{td-p}{(c)}]}, the boundary terms {\authorcorr turn out to be the external forces due to the confining walls and to have the} same legitimate status as the macroscopic observable force deriving from the physical potential $V(x)$.
With the specific purpose in mind to let {\authorcorr external-force} effects stand out, we have considered the {\authorcorr stationary-state} circumstance in which the initial wavefunction coincides with an eigenfunction [\REq{iwf}]; in such a simplified situation, the particle should be macroscopically observed resting at a position [\REq{mo.x.ss}] decided by the equilibrium between the macroscopic observable force due to the physical potential and the {\authorcorr external} forces [\REq{etc.m.p.ss}].
Ehrenfest's formulation [\REqq{pet}] does not account for the establishment of such an equilibrium of forces and would, therefore, raise a conflict in the relevant equations [\mbox{$\sigma=0$} in both \REq{etc.m.p.ss} and \REq{etc.m.p.ss.f}].
We have {\authorcorr complemented and} corroborated the theoretical findings with a numerical exercise in which the particle is an electric charge subjected to a uniform electrostatic field [\REq{pot.esf}].
We have solved numerically the nondimensional hamiltonian eigenvalue problem [\REq{tiSeq.pib.cef.nd}] governed by one main characteristic number [\REq{cn.a}] for the first four stationary states and obtained corresponding eigenvalues (\Rta{table}) and eigenfunctions (\Rfi{ef1-4.cef}) for three electric-field situations ($\alpha=0,10,100$).
We have illustrated the dependence of electric-charge macroscopic observable position (\Rfi{nmop}) and eigenvalues (\Rfi{nsev}) on electric field. 
The numerical results retrieved for the eigenfunctions' derivatives at the boundaries (\Rta{table}) prove beyond any doubt the indispensability [\mbox{$\sigma=1$} in \REq{etc.m.p.ss}, \REq{etc.m.p.ss.f} and \REq{etc.m.p.ss.f.nd}; see also \REq{tiSeq.pib.rs.iii.nd}] of the {\authorcorr external-force} presence in \REq{etc.m} and the incompleteness of Ehrenfest's theorem in the formulation he gave [\REqq{pet}].

{\authorcorr \section*{Acknowledgments}
We wish to thank B. Sutcliffe (Universit\'{e} Libre de Bruxelles) and S. De Vincenzo (Universidad Central de Venezuela, Caracas) for their availability to discuss aspects of the subject dealt with in this work and, above all, for their useful advises.
We also wish to thank A. R. Plastino (Universidad Naci\'{o}nal de Buenos Aires) for his insightful comments because they have broadened our perspective and, consequently, our understanding of the subject matter has definitely benefited by further thinking and elaboration induced by them.}

\vspace*{.5\baselineskip}
\section*{References}
\bibliographystyle{iopart-num}
     


\begin{thebibliography}{10}
\expandafter\ifx\csname url\endcsname\relax
  \def\url#1{{\tt #1}}\fi
\expandafter\ifx\csname urlprefix\endcsname\relax\def\urlprefix{URL }\fi
\providecommand{\eprint}[2][]{\url{#2}}

\bibitem{pe1927zfp}
Ehrenfest P 1927 {\em Zeitschrift f\"{u}r Physik\/} {\bf 45} 455--457

\bibitem{aer1928pr}
Ruark A~E 1928 {\em Physical Review\/} {\bf 31} 533--538

\bibitem{am11961}
Messiah A 1961 {\em Quantum mechanics\/} vol~1 (Amsterdam, The Netherlands:
  North-Holland Publishing Company)

\bibitem{cct1977}
Cohen-Tannoudji C, Diu B and Lalo\"{e} F 1977 {\em Quantum mechanics\/} vol~1
  (Wiley-VCH)

\bibitem{db1989}
Bohm D 1989 {\em Quantum theory\/} (New York NY: Dover Publications)

\bibitem{lb2000}
Ballentine L~E 2000 {\em Quantum mechanics\/} (Singapore: World Scientific
  Publishing Co,)

\bibitem{pa2005}
Atkins P and Friedman R 2005 {\em Molecular quantum mechanics\/} 4th ed
  (Oxford, UK: Oxford University Press)

\bibitem{dg2005}
Griffiths D~J 2005 {\em Introduction to quantum mechanics\/} 2nd ed (Upper Saddle River NJ: Pearson
  Education, Inc.)

\bibitem{dr1996ajp}
Rokhsar D~S 1996 {\em American Journal of Physics\/} {\bf 64} 1416--1418

\bibitem{cll2016arxiv}
Lin C~L 2016 Direct manifestation of {E}hrenfest's theorem in the infinite
  square well model arXiv:1609.07380
  \urlprefix\url{https://arxiv.org/abs/1609.07380}

\bibitem{rnh1973ajp}
Hill R~N 1973 {\em American Journal of Physics\/} {\bf 41} 736--738

\bibitem{nw1998me}
Wheeler N 1998 Ehrenfest's theorem ({M}iscellaneous Essays)\\
  \url{{\scriptsize
  https://www.reed.edu/physics/faculty/wheeler/documents/Quantum Mechanics/\mbox{Miscellaneous Essays/} \-Ehrenfest's Theorem.pdf}}

\bibitem{va2000inc}
Alonso V, {D}e Vincenzo S and Gonz\'{a}lez-{D}\'{i}az L 2000 {\em Il Nuovo
  Cimento B\/} {\bf 115} 155--164

\bibitem{va2001pla}
Alonso V, {D}e Vincenzo S and Gonz\'{a}lez-{D}\'{i}az L 2001 {\em Physics
  Letters A\/} {\bf 287} 23--30

{\authorcorr
\bibitem{sdv2013jop}
{{D}e Vincenzo } S 2013 {\em PRAMANA Journal of Physics\/} {\bf 80}
  797--810

\bibitem{sdv2013rbef}
{{D}e Vincenzo } S 2013 {\em Revista Brasileira de Ensino de
  F\'{i}sica\/} {\bf 35} (2308--)1--9
}

\bibitem{mj1989}
Jammer M 1989 {\em The conceptual development of quantum mechanics\/} ({\em The
  history of modern physics 1800 - 1950\/} vol~12) (Melville NY:American Institute of Physics)

\bibitem{lb1994pra}
Ballentine L~E, Yang Y and Zibin J~P 1994 {\em Physical Review A\/} {\bf 50}
  2854--2859

{\refereetwo 
\bibitem{rs1994}
{Shankar} R 1994 {\em Principles of Quantum Mechanics\/} 2nd ed
  (New York NY: Springer)
}

{\refereeone 
\bibitem{rp1979}
{Peierls} R 1979 {\em Surprises in Theoretical Physics\/} Princeton
  series in Physics (Princeton NJ: Princeton University Press)

\bibitem{jmc1982}
{Cassels} J~M 1982 {\em Basic Quantum Mechanics\/} 2nd ed (London:
  The Macmillan Press)

\bibitem{fr2009}
{Reif} F 2009 {\em Fundamentals of Statistical and Thermal
  Physics\/} (Long Grove IL: Waveland Press)
}

\bibitem{pd1967}
Dirac P~A~M 1967 {\em The principles of quantum mechanics\/} 4th ed The
  International Series of Monographs on Physics (Oxford, UK: Oxford University
  Press)

{\refereeone 
\bibitem{ldb1926cras}
{De Broglie} L 1926 {\em Comptes Rendus Hebdomadaires des
  S\'{e}ances de l'Acad\'{e}mie des Sciences\/} {\bf 183} 447--448

\bibitem{ldb1927crasjj}
{De Broglie} L 1927 {\em Comptes Rendus Hebdomadaires des
  S\'{e}ances de l'Acad\'{e}mie des Sciences\/} {\bf 184} 273--274

\bibitem{ldb1927crasjd}
{De Broglie} L 1927 {\em Comptes Rendus Hebdomadaires des
  S\'{e}ances de l'Acad\'{e}mie des Sciences\/} {\bf 185} 380--382

\bibitem{em1927zfp}
{Madelung} E 1927 {\em Zeitschrift f\"{u}r Physik\/} {\bf 40}
  332--326

\bibitem{db1952pr166}
{Bohm} D 1952 {\em Physical Review\/} {\bf 85} 166--179

\bibitem{db1952pr180}
{Bohm} D 1952 {\em Physical Review\/} {\bf 85} 180--193

\bibitem{tt1952ptp}
{Takabayasi} T 1952 {\em Progress of Theoretical Physics\/} {\bf
  8} 143--182

\bibitem{tt1953ptp}
{Takabayasi} T 1953 {\em Progress of Theoretical Physics\/} {\bf
  9} 187--222

\bibitem{hew1970prd}
{Wilhelm} H~E 1970 {\em Physical Review D\/} {\bf 1} 2278--2285

\bibitem{tt1983ptp}
{Takabayasi} T 1983 {\em Progress of Theoretical Physics\/} {\bf
  69} 1323--1344

\bibitem{pv2016f}
{Vadasz} P 2016 {\em Fluids\/} {\bf 1} 18/1--11

\bibitem{ln1971}
{Napolitano} L 1971 {\em Thermodynamique des syst\`{e}mes
  composites en \'{e}quilibre ou hors d'\'{e}quilibre\/} vol LXXI (Paris,
  France: Gauthier-Villars \'{E}diteur)

\bibitem{sdv2014rbef}
{{D}e Vincenzo } S 2014 {\em Revista Brasileira de Ensino de
  F\'{i}sica\/} {\bf 36} (2313--)1--5

\bibitem{bs1975}
{Sutcliffe} B 1975 Fundamentals of computational quantum chemistry
  {\em Computational Techniques in Quantum Chemistry and Molecular Physics\/}
  ({\em Proceedings of the NATO Advanced Study Institute held in Ramsau,
  Germany, 4-21 September 1974\/} vol~15) ed Diercksen G, Sutcliffe B and
  Veillard A (Dordrecht, The Netherlands: Reidel Publishing Company) pp 1--105
}

\bibitem{pw1935}
Pauling L and {Wilson} E~B 1935 {\em Introduction to quantum mechanics, with
  application to chemistry\/} (New York NY: McGraw-Hill)

\bibitem{pa2011jnaiam}
Amodio P and Settanni G 2011 {\em Journal of Numerical Analysis, Industrial and
  Applied Mathematics\/} {\bf 6} 1--13

\bibitem{pa2015cnsns}
Amodio P and Settanni G 2015 {\em Communications in Nonlinear Science and
  Numerical Simulation\/} {\bf 20} 641--649

\bibitem{pa2020lncs}
Amodio P and Settanni G 2020 Numerical strategies for solving multiparameter
  spectral problems {\em Proceedings of the 3rd Triennial International
  Conference and Summer School on Numerical Computations: Theory and
  Algorithms, NUMTA 2019, 15-21 June 2019, Crotone, Italy\/} ({\em Lecture
  Notes in Computer Science\/} vol 11974 LNCS) pp 298--305

{\authorcorr
\bibitem{dth1964}
ter Haar D 1964 {\em Selected problems in quantum mechanics\/} (London, UK:
  Infosearch Limited)
}

\end{thebibliography}
\providecommand{\newblock}{}

\end{document}